\documentclass[10pt,a4paper]{article}

\usepackage{arxiv}
\usepackage[utf8]{inputenc}
\usepackage[T1]{fontenc}
\usepackage{amsmath,amssymb,amsfonts}
\usepackage{graphicx}
\usepackage{booktabs}
\usepackage{xcolor}
\usepackage{url}
\usepackage{float}
\usepackage{placeins}
\usepackage[numbers,sort&compress]{natbib}
\usepackage{hyperref}

\graphicspath{{figures/}}

\title{Sparse data limit what a mechanistic corrosion model can predict}

\author{%
  Conrard Giresse Tetsassi Feugmo\\[2pt]
  \small Department of Chemistry and Department of Physics \& Astronomy,\\
  \small University of Waterloo, 200 University Avenue West,\\
  \small Waterloo, Ontario N2L 3G1, Canada\\[2pt]
  \small \href{mailto:cgtetsas@uwaterloo.ca}{cgtetsas@uwaterloo.ca}\\
  \small ORCID 0000-0002-8992-4335
}

\date{}

\begin{document}
\maketitle

\begin{abstract}
Mechanistic corrosion models are routinely fitted with four to
six parameters to a few measurements and extrapolated across service
lifetimes, yet whether the data determine them is rarely tested. We
identify a reduced point defect model for the duplex oxide on
Nb-stabilized AISI~347 in simulated
boiling-water-reactor water from published depth profiles at three
exposure times, by differentiable inversion graded against a
closed-form reference. Profile likelihood, a 1000-member bootstrap and
a prior-relaxation test agree that one of the five parameters is
undetermined and a second only weakly so, and that 95\% of the fit
statistic rests on three chromium points. The kinetics track an
empirical power law over the calibration window but diverge from it by
a factor of 3.5 to 6.3 at ten years, outside the propagated uncertainty
band. That separation, not the thickness it brackets, is what these
data determine; an exposure near 3000~hours would resolve it.
\end{abstract}


\section{Introduction}\label{sec:intro}

Computational corrosion models are built to predict degradation beyond
the conditions and durations that were measured. Whether they can do so
depends on a question asked far less often than the fit is reported: do
the calibration data determine the model's parameters, or merely fail to
contradict them? This paper puts that question to a mechanistic
passive-film model, on a dataset whose sparsity is typical rather than
exceptional, and follows the answer through to what it costs a
service-life prediction.

Nb-stabilized austenitic stainless steel AISI~347 (X6CrNiNb18-10) is
used in boiling-water-reactor (BWR) internals and piping for its
resistance to weld-decay sensitization, yet the evolution of its passive
oxide film under reactor-representative hydrothermal water has so far
had no mechanistic predictive model. Current practice for predicting
oxide behavior on light-water-reactor internals relies either on
operational water-chemistry criteria framed around threshold
electrochemical corrosion potential (ECP) and dissolved-oxygen
levels~\citep{macdonald1992hwc,lin1996ecp,yeh1995damage1,yeh1996damage2},
which say nothing about oxide thickness or composition, or on
data-driven machine-learning models of crack-growth
rate~\citep{igscc_ml_uq2025}, which are equally silent on the underlying
oxide chemistry.

The duplex oxide that austenitic stainless steels develop in
high-temperature water (a Cr-rich inner spinel layer growing by
solid-state transport beneath an Fe-rich outer layer formed by
precipitation) is well established for the 304/316
grades~\citep{robertson1991mechanism,stellwag1998mechanism,ziemniak2002corrosion,kuang2010oxidation,terachi2008corrosion,kim1999oxide,lister1987release},
and the point defect model (PDM) of Chao, Lin, and
Macdonald~\citep{chao1981pdm1,lin1981pdm2,macdonald1992pdm,macdonald2011history},
in the tradition of high-field oxidation
theory~\citep{cabrera1949oxidation}, provides the standard mechanistic
description: vacancy-mediated transport under a high interfacial field,
yielding closed-form thickness-versus-time laws tied to interfacial
reaction kinetics. Li et al.~\citep{li2020scwpdm} extended this
framework to supercritical water. For AISI~347 specifically, Veile et
al.~\citep{veile2024} recently provided the first detailed
characterization of the duplex oxide developed in simulated BWR
hydrothermal water, resolving Cr-rich barrier and Fe-rich outer layers
by energy-dispersive X-ray (EDX) depth profiling at three exposure
times; their analysis stops at empirical per-element power-law fits,
with no shared physical parameter set connecting the layers and no
assessment of how far those fits can be extrapolated.

This paper adapts the supercritical-water PDM
formulation~\citep{li2020scwpdm} to the subcritical hydrothermal-water
regime and identifies its kinetics from the Veile et al.\ dataset with a
differentiable, physics-informed inversion framework. The alloy
specificity of the result lies in the data, not the model structure:
the reaction network contains nothing Nb-specific, and what is delivered
is the first PDM identification against the only depth-profile dataset
available for this alloy and environment. The identification is the vehicle rather than the destination. The
question this paper is built around is a methodological gap that recurs
across four decades of PDM parameter estimation. Whether calibrated against electrochemical
impedance spectra~\citep{bojinov_mcm,ss316ln_pdm2024,ss321_pdm} or
against depth-profile data, four-to-six-parameter models are routinely
fit to three-to-six data points, and while parameter confidence
intervals and correlations are sometimes
reported~\citep{identifiability_electrochem1996}, we are not aware of a
formal profile-likelihood or bootstrap identifiability characterization
of a PDM fit. The distinction between ``the model fits'' and ``the data
constrain the parameters'' becomes consequential exactly when the fitted
model is meant to extrapolate beyond its calibration window, as here.
We close this gap with tools mature in adjacent
fields~\citep{raue2009identifiability,efron1979bootstrap,wieland2021identifiability},
standard in systems biology~\citep{biosystems_identifiability} and
increasingly applied to battery models~\citep{p2d_identifiability2021},
built directly into the inversion rather than bolted on afterward.

What that analysis returns, and what organizes the rest of this paper,
is a single answer arrived at from four directions. Of the five parameters the model needs, three exposure times determine three, though the third is sharp in profile and not under resampling; one is set by its prior and one only to within an order of magnitude.
Ninety-five percent of the fit statistic rests on the three chromium
points and eighty percent on one of them, and the residual pattern is
systematic in precisely the curvature that governs extrapolation. The spatial extension and the two candidate mobility models each fail an identifiability test for the same reason, and the nickel-zone closure fails on shape for it too.
The consequence is that the quantity these data determine is not the
ten-year thickness but the separation between the mechanistic and
empirical extrapolations (a factor of 3.5 to 6.3 at ten years), and that
no reanalysis of three exposures can sharpen it, while one designed
exposure of roughly 3000~hours would. Too few independent exposure times
is the paper's recurring finding, and each section below reaches it
independently.

Two results fall outside that thread and are stated here so they are not
mistaken for asides. Extended spatially with no additional free
parameters, the identified kinetics predict the measured composition
profile, and a structural argument shows why composition data cannot
constrain the defect diffusivity however many of them are taken. And
solving Poisson's equation together with the defect transport, rather
than imposing the field, shows that the constant-field closure on which
the PDM's field strength rests is not self-consistent at the defect
concentrations the model implies.

Those results are produced by a differentiable inversion framework
built on a physics-informed neural
solver~\citep{raissi2019pinn,karniadakis2021piml}, the neural
spectral-element method (NSEM)~\citep{feugmo2026nsem}. Kinetic parameters
are recovered by a hard-constrained inversion in which the trajectories
are generated by differentiable Runge--Kutta integration, so the
governing equations are satisfied exactly by construction; a classical
least-squares solver is an independent reference, and the two
agree to within 0.6\% on every parameter and to identical $\chi^2$. The
same machinery emulates the forward model with two neural backbones and,
in a soft-constrained joint field-and-parameter mode, extends to the
spatial transport problem. The model is exercised at three levels within
that one framework: zero-dimensional interfacial kinetics, one-dimensional
defect transport across the barrier layer, and a self-consistent solve of
the potential distribution together with that transport.

To be precise about what that machinery contributes: the
property doing the statistical work is \emph{differentiability} rather
than the neural representation: exact parameter gradients through a
vectorized solver are what make profile likelihood, a thousand-member
bootstrap, prior-relaxation scans and a dense operating-envelope grid
affordable, and closed-form models with automatic differentiation share
that property. The neural representation becomes necessary only where no
closed form or cheap forward solve exists, which is not the case here:
classical methods reproduce every result in this paper, and in the
spatial case more accurately. That is the point of the test case rather
than a defect of it: a framework can be graded only where an exact
reference exists, and grading it is the prerequisite for trusting it
where none does, a bound made explicit in Section~\ref{sec:discussion}. Our
prior work demonstrated this machinery on the PDM against synthetic
benchmarked data, cataloging four training-time failure
modes~\citep{farooqi2025pinnpdm}; the present paper is its first
application to real experimental data, and in carrying it out we
document and remedy a fifth, specific to sparse real data: joint
field-and-parameter training can settle into an equilibrium that
satisfies the data loss while leaving the governing equation only
approximately satisfied, exposed only by re-solving the recovered
parameters through an exact forward solve. Loss-balancing pathologies of
this general kind are a recognized theme in physics-informed
training~\citep{wang2022ntk,krishnapriyan2021failure}, and a
contemporaneous study reports a closely related diagnosis in an
unrelated domain~\citep{pinnverse2025}; the contribution here is its
documentation, its remedy, and a self-consistency check that travels
with the framework.

The paper is organized methods-first: Section~\ref{sec:methods} derives
the reduced model step by step and sets out the data, the statistical
machinery and the solver; Section~\ref{sec:results} carries the argument
above; and the Discussion reads the identified parameters
mechanistically before bounding how far that reading can be taken.

\section{Methods}\label{sec:methods}

\subsection{Reduced-model derivation}\label{sec:model-methods}

The model identified in this paper is a reduced form of the point defect
model in the formulation Li et al.~\citep{li2020scwpdm} developed for
supercritical water, adapted here to subcritical hydrothermal water.
Every parameter reported later is an output of the reduction rather than
an elementary rate constant, so the derivation is set out step by step
below and each symbol is defined where it first appears. The complete reaction network is given in
Table~\ref{tab:reactions}; the activated-complex derivation of the rate-constant
form, and known errata in the source
formulations~\citep{li2020scwpdm} are given in the Supplementary
Information (SI, Section~S1).

\begin{figure}[htbp]
\centering
\includegraphics[width=\textwidth]{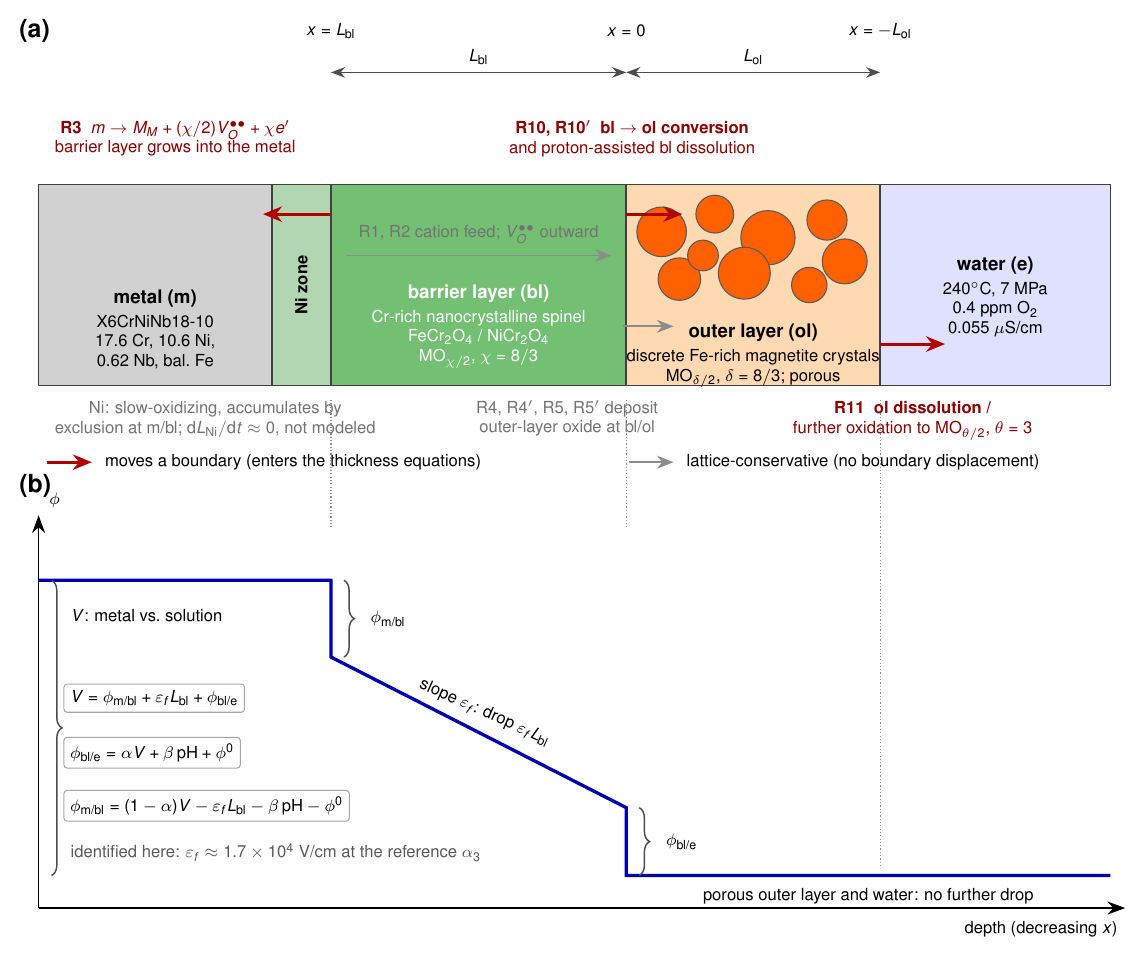}
\caption{Duplex-oxide model schematic, not to scale. (a) Layer
structure, drawn left to right in order of increasing distance from the
metal: metal $\vert$ Ni enrichment zone $\vert$ Cr-rich nanocrystalline
spinel barrier layer $\vert$ porous Fe-rich outer layer of discrete
magnetite crystals $\vert$ hydrothermal water. Interface positions carry
the coordinate convention of the SI (Section~S1.1), and the two state
variables $L_\text{bl}$ and $L_\text{ol}$ are dimensioned above the
stack. Each reaction is drawn at the interface where it acts, in red if
it displaces a boundary and therefore enters
Eqs.~\eqref{eq:odebl}--\eqref{eq:odeol}, in gray if it is
lattice-conservative: Reaction~3 grows the barrier layer into the metal,
Reactions~10 and $10'$ convert and dissolve it at the barrier-layer/outer-layer
interface, and Reaction~11 destroys the outer layer at the outer
surface. (b) Potential distribution across the same stack under the
constant-field postulate. The abscissa is aligned with (a) and the
ordinate is the electrostatic potential, so the decomposition $V =
\phi_\text{m/bl} + \varepsilon_f L_\text{bl} + \phi_\text{bl/e}$ reads
off the profile directly: a step at each of the two interfaces that
carries a drop, and between them the constant slope $\varepsilon_f$ of
Eq.~\eqref{eq:phimbl}. The porous outer layer carries no further drop.}
\label{fig:schematic}
\end{figure}

\textbf{Step 1: which reactions can move an interface.} The oxide Veile
et al.~\citep{veile2024} characterize on X6CrNiNb18-10 is a duplex scale
(Fig.~\ref{fig:schematic}a): a chromium-rich, nanocrystalline spinel
barrier layer of thickness $L_\text{bl}$ growing into the metal, and
discrete iron-rich magnetite crystals forming a porous outer layer of
thickness $L_\text{ol}$, with a nickel enrichment zone at the receding
metal/barrier-layer interface. The kinetics derived here treat that
nickel zone as an inert consistency check; a dedicated transport model
for it is developed in Section~\ref{sec:nickel}. Reactions occur at
three interfaces: metal/barrier layer (m/bl), barrier layer/outer layer
(bl/ol), and outer layer/water. Of the fourteen reactions in the network
(eleven numbered, three of them with primed variants), only four
displace a boundary. Reaction~3, $m \rightarrow M_M +
(\chi/2)\,V_O^{\bullet\bullet} + \chi e'$, in which a metal atom $m$
enters the oxide as a lattice cation $M_M$ while generating oxygen
vacancies $V_O^{\bullet\bullet}$ (charge number $\chi$, defined in
Step~3) and electrons $e'$, grows the barrier layer into the metal;
Reactions~10 and $10'$ convert or dissolve barrier-layer oxide;
Reaction~11 dissolves or further oxidizes outer-layer oxide. The other
ten are lattice-conservative: they carry cations across the barrier
layer and deposit them in the outer layer without themselves moving a
boundary. The thickness equations therefore contain only these four
rates.

\begin{table}[htbp]
\centering
\caption{The complete interfacial reaction network of the reduced point
defect model, retained from the supercritical-water
formulation~\citep{li2020scwpdm}. $m$ is a metal atom, $M_M$ a lattice
cation, $V_M^{\chi\prime}$ a cation vacancy, $M_i^{\chi+}$ a cation
interstitial, $V_O^{\bullet\bullet}$ an oxygen vacancy, $O_O$ a lattice
oxygen, $v_m$ a vacancy in the metal lattice and $e'$ an electron;
$\chi$ and $\delta$ are the cation valences in the barrier and outer
layers and $\theta$ the valence in the dissolved or further-oxidized
product. Only the four reactions in bold displace a boundary and so
enter Eqs.~\eqref{eq:blraw} and \eqref{eq:olraw}; the other ten are
lattice-conservative, and the cation fluxes they carry are eliminated by
the constant-volume closure of Step~6. Primed variants are alternative
routes to the same defect change, favored in condensed water.}
\label{tab:reactions}
\footnotesize
\setlength{\tabcolsep}{4pt}
\begin{tabular}{@{}c@{\ \ }l@{\quad}p{0.30\textwidth}@{}}
\toprule
\# & Reaction & Role \\
\midrule
\multicolumn{3}{@{}l}{\emph{metal/barrier-layer interface, $x = L_\text{bl}$}}\\
1 & $m + V_M^{\chi\prime} \rightarrow M_M + v_m + \chi e'$
  & annihilates a cation vacancy; outer-layer cation feed, vacancy path \\
2 & $m \rightarrow M_i^{\chi+} + v_m + \chi e'$
  & generates a cation interstitial; outer-layer cation feed, interstitial path \\
3 & $m \rightarrow M_M + (\chi/2)\, V_O^{\bullet\bullet} + \chi e'$
  & generates oxygen vacancies; \textbf{grows the barrier layer into the metal} \\[3pt]
\multicolumn{3}{@{}l}{\emph{barrier-layer/outer-layer interface, $x = 0$}}\\
4 & $M_M + (\delta/4)\,\mathrm{O_2} + \chi e' \rightarrow
     V_M^{\chi\prime} + \mathrm{MO}_{\delta/2}$
  & generates a cation vacancy; deposits outer layer, direct $\mathrm{O_2}$ route \\
$4'$ & $M_M \rightarrow V_M^{\chi\prime} + M^{\delta+} + (\delta-\chi)\, e'$
  & generates a cation vacancy; cation ejected to solution \\
5 & $M_i^{\chi+} + (\delta/4)\,\mathrm{O_2} + \chi e' \rightarrow
     \mathrm{MO}_{\delta/2}$
  & annihilates an interstitial; deposits outer layer \\
$5'$ & $M_i^{\chi+} \rightarrow M^{\delta+} + (\delta-\chi)\, e'$
  & annihilates an interstitial; cation ejected to solution \\
6 & $V_O^{\bullet\bullet} + \tfrac12 \mathrm{O_2} + 2e' \rightarrow O_O$
  & annihilates an oxygen vacancy, $\mathrm{O_2}$ route \\
7 & $V_O^{\bullet\bullet} + \mathrm{H_2O} + 2e' \rightarrow O_O + \mathrm{H_2}$
  & annihilates an oxygen vacancy, water route \\
8 & $V_O^{\bullet\bullet} + \mathrm{H_2O} \rightarrow O_O + 2\mathrm{H^+}$
  & annihilates an oxygen vacancy with proton release, classical PDM \\
9 & $\mathrm{H_2O} + 2e' \rightarrow \tfrac12 \mathrm{H_2} + \mathrm{OH^-}$
  & cathodic partial reaction; carries no defect \\
10 & $\mathrm{MO}_{\chi/2} + \frac{\delta-\chi}{4}\,\mathrm{O_2}
      \rightarrow \mathrm{MO}_{\delta/2}$
  & \textbf{converts barrier oxide to outer oxide}, chemical route \\
$10'$ & $\mathrm{MO}_{\chi/2} + \chi\,\mathrm{H^+} \rightarrow
        M^{\delta+} + (\chi/2)\,\mathrm{H_2O} + (\delta-\chi)\, e'$
  & \textbf{proton-assisted barrier-layer dissolution} \\[3pt]
\multicolumn{3}{@{}l}{\emph{outer-layer/water interface, $x = -L_\text{ol}$}}\\
11 & $\mathrm{MO}_{\delta/2} + \frac{\theta-\delta}{4}\,\mathrm{O_2}
      \rightarrow \mathrm{MO}_{\theta/2}(d)$
  & \textbf{destroys the outer layer}: dissolution or further oxidation \\
\bottomrule
\end{tabular}
\end{table}

\textbf{Step 2: the potential distribution and the constant-field
postulate.} Let $V$ be the potential difference between metal and
solution and $\phi_\text{m/bl}$, $\phi_\text{bl/e}$ the drops at the
metal/barrier-layer and barrier-layer/water interfaces. The PDM
postulates that the electric field inside the barrier layer is constant
at $\varepsilon_f$, independent of both $V$ and $L_\text{bl}$, and that
the outer drop is linear in potential and pH,
\begin{equation}
\phi_\text{bl/e} = \alpha V + \beta\,\mathrm{pH} + \phi^0,
\label{eq:phible}
\end{equation}
where $\alpha = \partial\phi_\text{bl/e}/\partial V$ is the
polarizability of that interface, $\beta =
\partial\phi_\text{bl/e}/\partial\,\mathrm{pH} < 0$, and $\phi^0$ is the
drop at $V = \mathrm{pH} = L_\text{bl} = 0$. Summing the drops across
the stack, $V = \phi_\text{m/bl} + \varepsilon_f L_\text{bl} +
\phi_\text{bl/e}$, so that
\begin{equation}
\phi_\text{m/bl} = (1-\alpha)V - \varepsilon_f L_\text{bl}
- \beta\,\mathrm{pH} - \phi^0 .
\label{eq:phimbl}
\end{equation}
The term that carries the kinetics is $-\varepsilon_f L_\text{bl}$: the
potential available to drive the metal/barrier-layer reactions falls
linearly as the film thickens.

\textbf{Step 3: the rate constants inherit that thickness dependence.}
Activated-complex theory with partial charge transfer gives every
interfacial rate constant the form
\begin{equation}
k_i = k_i^{0}\, e^{\,a_i V}\, e^{\,b_i L_\text{bl}}\, e^{\,c_i\,\mathrm{pH}},
\label{eq:rateconstant}
\end{equation}
with $k_i^0$ the standard rate constant of Reaction~$i$ and $a_i$,
$b_i$, $c_i$ fixed by which interfacial drop that reaction sees (SI, Table~S2). Reactions at bl/ol and at the outer surface see
$\phi_\text{bl/e}$, which by Eq.~\eqref{eq:phible} carries no
$L_\text{bl}$ dependence, so $b_i = 0$ for all of them. The m/bl
reactions see $\phi_\text{m/bl}$ and therefore inherit
Eq.~\eqref{eq:phimbl}; for the growth reaction this gives
\begin{equation}
b_3 = -\alpha_3 \chi \gamma \varepsilon_f < 0 ,
\label{eq:b3}
\end{equation}
in which $\alpha_3$ is the transfer coefficient of Reaction~3
(dimensionless), $\chi = 8/3$ the spinel-averaged cation charge number,
and $\gamma = F/RT = 22.6$~V$^{-1}$ at 240$^\circ$C, with $F$ the
Faraday constant, $R$ the gas constant and $T$ the absolute temperature.
The negative sign is the kinetic content of the model: growth
decelerates exponentially in the film's own thickness, which is what
produces direct-logarithmic kinetics and, whenever a destruction term is
nonzero, a finite steady-state thickness.

\textbf{Step 4: the barrier-layer thickness equation.} Only the
non-conservative reactions of Step~1 move the barrier layer's
boundaries, so its thickness obeys
\begin{equation}
\frac{\mathrm{d}L_\text{bl}}{\mathrm{d}t}
= \Omega_\text{bl}\left(k_3 - k_d\right)
\frac{\rho^0_\text{bl}}{\rho_\text{bl}},
\qquad
k_d := k_{10} C_\mathrm{O}^{\,q}
+ k_{10'}\left(C_{\mathrm{H}^+}/C^0_{\mathrm{H}^+}\right)^{s},
\label{eq:blraw}
\end{equation}
where $\Omega_\text{bl}$ is the molar volume of the barrier oxide per
cation and $\rho^0_\text{bl}/\rho_\text{bl}$ a porosity correction
(set to unity for both layers throughout, for the reason given in
Step~6),
$C_\mathrm{O}$ the dissolved-oxygen concentration,
$C_{\mathrm{H}^+}/C^0_{\mathrm{H}^+}$ the proton activity ratio, and
$q$ and $s$ the corresponding reaction orders. The lumped rate $k_d$ is
the total barrier-layer destruction rate. Substituting
Eq.~\eqref{eq:rateconstant} for $k_3$ at fixed $V$ and pH puts
Eq.~\eqref{eq:blraw} in the two-parameter working form used throughout,
with
\begin{equation}
A_\text{bl} = \Omega_\text{bl}\, k_3^{0}\, e^{\,a_3 V}\,
e^{\,c_3\,\mathrm{pH}}\, \frac{\rho^0_\text{bl}}{\rho_\text{bl}},
\qquad
C_\text{bl} = \Omega_\text{bl}\, k_d\,
\frac{\rho^0_\text{bl}}{\rho_\text{bl}} .
\label{eq:AblCbl}
\end{equation}

\textbf{Step 5: the outer layer before coupling.} The outer layer grows
from cations transmitted through the barrier layer, from barrier oxide
destroyed by Reactions~10/$10'$ and re-precipitated as outer oxide, and
shrinks by dissolution:
\begin{equation}
\frac{\mathrm{d}L_\text{ol}}{\mathrm{d}t}
= \Omega_\text{ol}\left(k_1 C_{V_M}^{L} + k_2 + k_d
- k_{11} C_\mathrm{O}^{\,r}\right)
\frac{\rho^0_\text{ol}}{\rho_\text{ol}},
\label{eq:olraw}
\end{equation}
in which $\Omega_\text{ol}$ and $\rho^0_\text{ol}/\rho_\text{ol}$ are
the outer layer's molar volume per cation and porosity correction,
$k_1 C_{V_M}^{L}$ and $k_2$ are the cation fluxes arriving by the
vacancy and interstitial paths respectively ($C_{V_M}^{L}$ being the
cation-vacancy concentration at the m/bl interface), and
$k_{11} C_\mathrm{O}^{\,r}$ is the outer-layer dissolution rate with
order $r$. Following Ref.~\citep{li2020scwpdm}, the whole of $k_d$ is
credited to the outer layer, which slightly overstates its supply
because the proton-assisted route $k_{10'}$ ejects cations into solution
rather than into oxide; in this ultrapure-water environment $k_d$ is
fitted to zero in any case. Written this way the two layers are still independent, and
the six rates in Eqs.~\eqref{eq:blraw} and \eqref{eq:olraw} are far more
than three exposure times can determine.

\textbf{Step 6: the constant-volume constraint collapses the coupling.}
Marker experiments place the bl/ol interface at the original metal
surface, so the oxide grows into the metal rather than outward from it:
metal is consumed at exactly the rate the barrier layer advances,
$\mathrm{d}L_m/\mathrm{d}t = \mathrm{d}L_\text{bl}/\mathrm{d}t$, where
$L_m$ is the thickness of metal consumed,
$\mathrm{d}L_m/\mathrm{d}t = \Omega_m (k_1 C_{V_M}^{L} + k_2 + k_3)$ and
$\Omega_m$ is the metal's molar volume per atom. Equating the two and using
Eq.~\eqref{eq:blraw} gives the constant-volume condition
\begin{equation}
k_1 C_{V_M}^{L} + k_2 + k_3 = \mathrm{PBR}\,(k_3 - k_d),
\qquad
\mathrm{PBR} := \Omega_\text{bl}/\Omega_m ,
\label{eq:cv}
\end{equation}
$\mathrm{PBR}$ being the Pilling--Bedworth ratio of the barrier oxide.
The metal is dense, so the left-hand side of Eq.~\eqref{eq:cv} carries no
porosity correction while the right-hand side inherits
$\rho^0_\text{bl}/\rho_\text{bl}$ from Eq.~\eqref{eq:blraw}. Writing
Eq.~\eqref{eq:cv} without that factor is therefore the dense-oxide
convention $\rho^0_\text{bl}/\rho_\text{bl} =
\rho^0_\text{ol}/\rho_\text{ol} = 1$, and equal porosity between the two
layers is not sufficient for it: carrying both corrections through the
substitution below leaves the outer-to-barrier coefficient as
$(\Omega_\text{ol}/\Omega_\text{bl})(\rho^0_\text{ol}/\rho_\text{ol})
(\mathrm{PBR} - \rho_\text{bl}/\rho^0_\text{bl})$, which collapses to the
form used below only when both are unity. The convention bears on the
physical quantities recovered in Step~9, not on the fitted parameters,
which absorb it.
Substituting Eq.~\eqref{eq:cv} into Eq.~\eqref{eq:olraw}, the two
destruction terms combine exactly, $(\mathrm{PBR}-1)k_3 -
\mathrm{PBR}\,k_d + k_d = (\mathrm{PBR}-1)(k_3-k_d)$, leaving the
cation fluxes eliminated and the outer-layer growth rate an exact affine
function of the barrier-layer growth rate,
\begin{equation}
\frac{\mathrm{d}L_\text{ol}}{\mathrm{d}t}
= (\mathrm{PBR}-1)\frac{\Omega_\text{ol}}{\Omega_\text{bl}}
\frac{\mathrm{d}L_\text{bl}}{\mathrm{d}t}
- \Omega_\text{ol}\, k_{11} C_\mathrm{O}^{\,r}\,
\frac{\rho^0_\text{ol}}{\rho_\text{ol}} .
\label{eq:olcoupled}
\end{equation}
This is the step that ties the two layers to one growth law, and it is
what makes the mechanistic and empirical descriptions extrapolate
differently in Section~\ref{sec:predictions}.

\textbf{Step 7: reduction to apparent parameters.} At open circuit $V$
and pH are constant over an exposure (the water is neutral, unbuffered
and of very low conductivity; the neutral pH at 240$^\circ$C is
approximately 5.6), so the factor $e^{\,a_3 V + c_3\,\mathrm{pH}}$ in
Eq.~\eqref{eq:AblCbl} is a constant multiplier that thickness-versus-time
data cannot separate from $k_3^0$. The same is true of the molar volumes
and porosity corrections. What the data can determine is therefore not
the elementary rate constants but the lumped groups of
Eqs.~\eqref{eq:AblCbl}, \eqref{eq:cv} and \eqref{eq:olcoupled}, giving
the apparent-parameter system fitted in this paper:
\begin{align}
\frac{\mathrm{d}L_\text{bl}}{\mathrm{d}t} &=
A_\text{bl}\, e^{\,b_3 L_\text{bl}} - C_\text{bl},
& L_\text{bl}(0) &= L_0,
\label{eq:odebl}\\
\frac{\mathrm{d}L_\text{ol}}{\mathrm{d}t} &=
\mathrm{PBR}_\text{eff}\,
\frac{\mathrm{d}L_\text{bl}}{\mathrm{d}t} + C_x,
& L_\text{ol}(0) &= L_\text{ol,0},
\label{eq:odeol}
\end{align}
where $t$ is exposure time (h) and $L_\text{bl}(t)$, $L_\text{ol}(t)$
are the two layer thicknesses (nm). The five rate parameters are
$A_\text{bl}>0$ (nm/h), the apparent growth prefactor of
Eq.~\eqref{eq:AblCbl}, which absorbs the potential- and pH-dependent
part of the growth rate constant; $b_3<0$ (nm$^{-1}$), the apparent
field-related exponent of Eq.~\eqref{eq:b3}, which sets how steeply
growth decelerates as the film thickens; $C_\text{bl}\ge0$
(nm/h), the barrier-layer destruction rate;
\begin{equation}
\begin{aligned}
\mathrm{PBR}_\text{eff} &:= (\mathrm{PBR}-1)\,
\frac{\Omega_\text{ol}}{\Omega_\text{bl}}
&&\text{(dimensionless)},\\[2pt]
C_x &:= -\Omega_\text{ol}\, k_{11} C_\mathrm{O}^{\,r}\,
\frac{\rho^0_\text{ol}}{\rho_\text{ol}} \;\le\; 0
&&(\mathrm{nm/h}),
\end{aligned}
\label{eq:pbreffCx}
\end{equation}
the apparent outer-to-barrier growth-rate ratio and the outer-layer
dissolution term, the latter written with a sign convention such that a
negative value removes outer-layer material. The initial conditions
$L_\text{bl}(0)=L_0$ and $L_\text{ol}(0)=L_\text{ol,0}$ (both nm)
complete the seven-parameter set.

\textbf{Step 8: closed-form solutions.} Both equations integrate
analytically at constant parameters. The barrier layer gives
\begin{equation}
L_\text{bl}(t) = L_0 - \frac{1}{b_3}
\ln\!\left[1 + \frac{A_\text{bl}}{C_\text{bl}}\, e^{\,b_3 L_0}
\left(e^{-b_3 C_\text{bl} t} - 1\right)\right] - C_\text{bl}\, t ,
\label{eq:blclosed}
\end{equation}
which for $C_\text{bl}\to0$ reduces to the direct-logarithmic law
$L_\text{bl}(t) = L_0 - b_3^{-1}\ln(1 - b_3 A_\text{bl} e^{\,b_3 L_0} t)$
and for $C_\text{bl}>0$ saturates at a finite
$L_\text{bl,ss}$; those two branches are the M4 and M5 variants compared
in Section~\ref{sec:predictions}. Because
Eq.~\eqref{eq:odeol} is affine in
$\mathrm{d}L_\text{bl}/\mathrm{d}t$, the outer layer follows by direct
integration with no further approximation,
\begin{equation}
L_\text{ol}(t) = L_\text{ol,0}
+ \mathrm{PBR}_\text{eff}\left(L_\text{bl}(t) - L_0\right) + C_x\, t .
\label{eq:olclosed}
\end{equation}
Equations~\eqref{eq:blclosed}--\eqref{eq:olclosed} are evaluated in log
space so that they remain numerically exact in both the
$|b_3|L\ll1$ and $|b_3|L=O(1)$ limits (SI, Section~S1.7); this
closed form, not a numerical integration, is what the fits and the
bootstrap evaluate, which is why a thousand refits are affordable.

\textbf{Step 9: recovering physical quantities.} The fitted apparent
parameters map back to the underlying physics through
\begin{equation}
\varepsilon_f = \frac{-b_3}{\alpha_3 \chi \gamma},
\qquad
\mathrm{PBR} = 1 + \mathrm{PBR}_\text{eff}\,
\frac{\Omega_\text{bl}}{\Omega_\text{ol}},
\qquad
k_d = \frac{C_\text{bl}}{\Omega_\text{bl}} ,
\label{eq:recover}
\end{equation}
each inverting one of Eqs.~\eqref{eq:b3}, \eqref{eq:pbreffCx} and \eqref{eq:AblCbl}. The second and third both inherit the dense-oxide convention of Step~6, the second because Eq.~\eqref{eq:olcoupled} is derived under it and the third because the porosity correction does not cancel out of Eq.~\eqref{eq:AblCbl}; the first involves no porosity correction and needs neither. The
first is evaluated with $b_3$ converted to the cgs units in which field
strengths are conventionally quoted
($1$~nm$^{-1} = 10^{7}$~cm$^{-1}$), so that
$b_3 = -0.0125$~nm$^{-1} = -1.25\times10^{5}$~cm$^{-1}$ gives
$\varepsilon_f$ in V\,cm$^{-1}$. Only the first requires an assumption
the thickness data cannot supply, the transfer coefficient $\alpha_3$, which is why
the field strength is reported over a range of $\alpha_3$ in
Section~\ref{sec:discussion} and why the identification is stated in terms of
$b_3$ rather than $\varepsilon_f$.

Two independent checks establish that the closed-form solutions are
exact solutions of the reduced system; the exactness of the reduction
itself rests on the constant-volume closure of Step~6 and the
equal-porosity convention. The closed-form solutions
\eqref{eq:blclosed}--\eqref{eq:olclosed} agree with Radau numerical
integration of Eqs.~\eqref{eq:odebl}--\eqref{eq:odeol} to better than
$6\times10^{-11}$~nm (SI, Fig.~S1), and the affine outer-layer form
\eqref{eq:olclosed} agrees with the algebraically heavier closed form of
Ref.~\citep{li2020scwpdm} to better than $1\times10^{-10}$~nm.

\subsection{Experimental data}\label{sec:data}

The reduction leaves seven apparent parameters, and everything that can
be asked of them is set by the one dataset that exists for this alloy in
this environment. Veile et al.~\citep{veile2024} exposed X6CrNiNb18-10 coupons to
simulated BWR hydrothermal water (240$^\circ$C, 7~MPa, 0.4~ppm dissolved
O$_2$) for 72, 168, and 480~h and measured element-resolved oxide layer
thicknesses by focused-ion-beam cross-sections with EDX line scans.
Table~\ref{tab:data} reproduces the digitized layer-thickness means used
as the inversion target throughout this paper. The digitization
was checked against their own reported empirical fits: a scan-level
linear-space least-squares refit of $L = k\,t^n$ to our digitized
chromium data reproduces their published coefficients to four
significant figures ($k = 6.522$ versus their 6.521, $n = 0.4964$
versus 0.4964), which bounds the digitization error well below the
scan-to-scan scatter. For the iron layer the reported scan-to-scan
physical scatter of the discrete-crystal outer layer exceeds
digitization precision, so their reported layer means are used as the
authoritative data. No synthetic or simulated data enters the
kinetic identification; synthetic data appears only as an explicitly
labeled recovery test (Section~\ref{sec:pinn}).

\subsection{Statistical inversion and identifiability analysis}\label{sec:stats}

Six layer means at three exposure times against a five-parameter model
is the ratio that dictates both the objective below and the
identifiability machinery carried alongside it: at this sample size a
good fit is not evidence that the parameters are determined, and the
two questions have to be asked separately. All fits minimize the
weighted least-squares objective
\begin{equation}
\begin{split}
\Phi(\theta) = {}&
\sum_{i=1}^{3} \frac{\left[L_\text{bl}(t_i;\theta) - \bar{L}_{\text{Cr},i}\right]^2}{\sigma_{\text{Cr},i}^2}
+ \sum_{i=1}^{3} \frac{\left[L_\text{ol}(t_i;\theta) - \bar{L}_{\text{Fe},i}\right]^2}{\sigma_{\text{Fe},i}^2} \\
&+ \left(\frac{\ln L_0 - \ln 2\,\text{nm}}{0.60}\right)^{\!2},
\end{split}
\label{eq:objective}
\end{equation}
where $\theta$ is the parameter vector being estimated, the index $i$
runs over the three exposure times $t_i \in \{72, 168, 480\}$~h,
$L_\text{bl}(t_i;\theta)$ and $L_\text{ol}(t_i;\theta)$ are the
closed-form model thicknesses at those times,
$\bar{L}_{\text{Cr},i}$ and $\bar{L}_{\text{Fe},i}$ are the measured
layer means and $\sigma_{\text{Cr},i}$, $\sigma_{\text{Fe},i}$ their
standard errors (Table~\ref{tab:data}), and the final term is a
Gaussian prior in log-parameter space on the initial barrier-layer
thickness $L_0$, centered on the 1--5~nm air-formed film on high-Cr
stainless steel with a log-space width $\sigma_{\ln L_0} = 0.60$. It is the only
prior in the objective, and it is carried because the exposure-time data
give $L_0$ no interior optimum at all: without it the estimate runs to
zero (Section~\ref{sec:identifiability}). In particular no prior is placed on
$\mathrm{PBR}_\text{eff}$, which is determined by the likelihood alone
throughout; the stoichiometric reasoning that decides what that value
can be compared against is taken up in Section~\ref{sec:discussion}.
The classical minimization
uses Levenberg--Marquardt least squares in log-parameter space (which
enforces positivity by construction), warm-started and verified against
grid restarts; convergence tolerances are $10^{-8}$. The equivalent
hard-mode physics-informed inversion (Section~\ref{sec:nsem}) minimizes
the same objective with the trajectories generated by differentiable
Runge--Kutta integration. The two are independent routes to the same
optimum rather than a primary and a check, and Section~\ref{sec:pinn}
reports how closely they agree; the deterministic parameter values and
curves given throughout
(Figs.~\ref{fig:fits}--\ref{fig:longtime}) are the converged optimum
common to both.

Profile-likelihood scans fix one parameter on a grid of $\pm1.5$
natural-log units about the optimum (61 points) and re-optimize all
others under the full objective~\eqref{eq:objective}; the
$\Delta\chi^2$ values reported are computed over the data residuals
alone, so a prior-dominated parameter is exposed as flat rather than
masked. The parametric bootstrap resamples each layer mean from a
Gaussian with its standard error (1000 members, all refits converged;
positivity is guaranteed by the log-space parametrization) and refits
the full objective.

Because one parameter is prior-set and a second only weakly determined,
the number of data-determined parameters is between three and five, and the residual
degrees of freedom for the accepted five-parameter model are
correspondingly between 1 and 3 for the six data points. Goodness-of-fit
statements are therefore given for both endpoints of this
bracket rather than a single nominal degree-of-freedom count.

\subsection{Physics-informed spectral-element solver}\label{sec:nsem}

\begin{figure}[htbp]
\centering
\includegraphics[width=\textwidth]{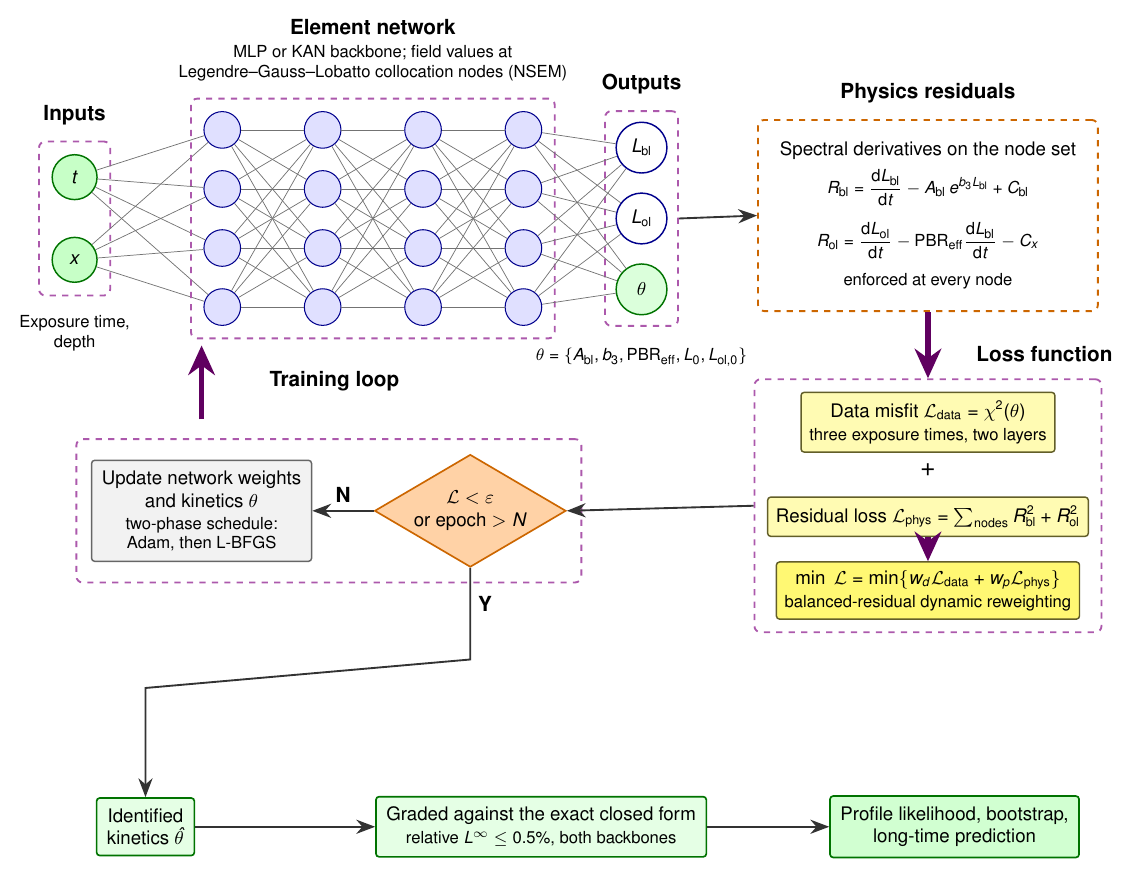}
\caption{How the reduced model is inverted. Exposure time, and for the
spatial extension depth, enters an element network whose field values
are evaluated at Legendre--Gauss--Lobatto collocation nodes; the
network returns the two layer thicknesses and the five kinetic
parameters $\theta$. Spectral differentiation on the same node set
gives the residuals of Eqs.~\eqref{eq:odebl}--\eqref{eq:odeol}, which
are enforced at every node and combined with the data misfit of
Eq.~\eqref{eq:objective} by a balanced-residual dynamic reweighting of
the two terms; a two-phase schedule, Adam then L-BFGS, drives the total
loss to convergence. The identified kinetics are graded against the
exact closed form of Eqs.~\eqref{eq:blclosed}--\eqref{eq:olclosed}
before they are used for profile likelihood, the bootstrap, and the
long-time prediction. The solver is the neural spectral-element method
(NSEM)~\citep{feugmo2026nsem}; architecture and hyperparameters for
each configuration are tabulated in the SI (Section~S4).}
\label{fig:nsem_workflow}
\end{figure}

The engine that minimizes that objective, and the forward solver used
for the spatial extension later, are the same piece of machinery,
described here once. The physics-informed forward and inverse solves use
the neural
spectral-element method (NSEM)~\citep{feugmo2026nsem}, in which each
field is evaluated only at fixed Legendre--Gauss--Lobatto quadrature
nodes and derivatives are taken with precomputed spectral
differentiation matrices, giving a deterministic loss that L-BFGS can
drive to residuals well below the accuracy floor of random-collocation
training. The solver was applied to the PDM family in our prior
work~\citep{farooqi2025pinnpdm}. Each field is represented by an
element network evaluated on a collocation node set, physical residuals
are enforced at the nodes through a first-integral or collocation
formulation, loss terms are combined by a balanced-residual dynamic
reweighting aggregator, and optimization proceeds in two phases (Adam,
then L-BFGS). Full architecture, loss definitions, and hyperparameters
for every configuration are tabulated in the SI (Section~S4), which also
summarizes the workflow in one diagram (Fig.~\ref{fig:nsem_workflow}).

Three configurations produce the results reported here; a fourth, a
pure forward solve, is used only for the backbone verification
described below. The three are: a ``hard'' zero-dimensional inverse, in
which the kinetic parameters are the only learnable quantities and the
trajectories are generated by differentiable Runge--Kutta integration,
so Eqs.~\eqref{eq:odebl}--\eqref{eq:odeol} are satisfied exactly by
construction; a ``soft'' zero-dimensional inverse, in which field
networks represent the trajectories directly and are trained jointly
with the parameters against a combined data-and-physics loss (the
configuration in which the failure mode of Section~\ref{sec:pinn} is
exposed); and the spatial extension of Section~\ref{sec:spatial}, where
no differentiable-integrator shortcut exists and the soft-mode field
representation is the only option, verified against an independent
Newton solve on the same node set.

Every configuration is checked against an independent reference before
its results are used. The classical baseline itself is established
to machine precision (Section~\ref{sec:model-methods}). Two
neural backbones, a multilayer perceptron (MLP) and a
Kolmogorov--Arnold network (KAN)~\citep{liu2024kan}, are evaluated in
the same harness to test backbone sensitivity, both to the same
acceptance criterion (relative $L^\infty$ error $\le0.5\%$ against the
closed form). All results are backed by an automated test suite (25
tests) checking parity between closed-form, Newton, and neural
solutions; the suite ships with the code release (see Code
availability).

\section{Results}\label{sec:results}

The results follow the order the argument requires. We first establish
that the reduced model of Section~\ref{sec:model-methods} can be
identified from these data at all, and which of its variants the data
select (Section~\ref{sec:ladder}). We then ask the question that decides
what the identification is worth, namely which parameters the three
exposure times actually determine
(Section~\ref{sec:identifiability}), and follow it straight to what
those limits cost the prediction the model is fitted to make
(Section~\ref{sec:predictions}). The remaining sections spend the
identified kinetics: on a spatial extension that predicts composition
with no further free parameters (Section~\ref{sec:spatial}), and on the
one feature of the scale those kinetics fail to describe
(Section~\ref{sec:nickel}). We close by grading the inversion framework
itself against the exact reference this problem happens to admit, which
is also where the failure mode we document is exposed
(Section~\ref{sec:pinn}).

\subsection{Differentiable inversion and the model ladder}\label{sec:ladder}

We identify the reduced kinetics by minimizing a weighted least-squares
objective (Section~\ref{sec:stats}, Eq.~\eqref{eq:objective}), solving
the inverse problem two independent ways: a hard-constrained
physics-informed neural inversion, in which the trajectories are produced
by differentiable Runge--Kutta integration so that
Eqs.~\eqref{eq:odebl}--\eqref{eq:odeol} hold exactly, and a classical
Levenberg--Marquardt solver. Section~\ref{sec:pinn} grades the two
against each other; they agree closely enough that a single identified
parameter set is reported, and the ladder below is used only to decide
model structure.

We fit the reduced system~\eqref{eq:odebl}--\eqref{eq:odeol} to the
joint Cr and Fe thickness means (Table~\ref{tab:data}) in a nested model
ladder
(Table~\ref{tab:ladder}), each variant adding degrees of freedom to
test whether the flexibility is supported by the data. The
four-parameter core model (M1, $C_\text{bl}=C_x=0$,
$L_\text{ol,0}=0$) reaches $\chi^2 = 7.15$ only by driving
$\mathrm{PBR}_\text{eff}$ to 2.43, and still cannot reproduce the
disproportionately thick early iron layer under strict constant-volume
coupling from zero initial thickness; adding a free dissolution rate
(M2) does not help, because the fit drives that rate to zero. Releasing
the coupling entirely in the six-parameter M3 variant does not produce a
better model but a degenerate one, and with no prior to hold it back the
degeneracy is now the interior optimum rather than something a profile
has to expose: the growth prefactor and dissolution rate diverge
together to $\sim\!10^{3}$~nm/h while the field parameter
collapses by four orders of magnitude to
$-4.4\times10^{-6}$~nm$^{-1}$, buying the lowest data $\chi^2$ in the
ladder (4.11) through an unphysical linear-growth escape route. Its
ten-year barrier-layer prediction is 155~nm against M4's 535~nm, so the
$\chi^2$ ordering and the physical ordering disagree outright. That M3
wins on fit and loses on physics is the clearest statement in this paper
of why $\chi^2$ cannot by itself select a model on six data points.
Parameter correlations of this kind have long been discussed
qualitatively in PDM fitting; the explicit demonstration of the full
ridge, its escape direction, and its capture of the unconstrained
optimum on real PDM data is, to our knowledge, new.

The trap is resolved by a single additional parameter, the initial
outer-layer thickness (M4): a nonzero starting population of Fe-rich
outer-layer crystals captures the rapid initial precipitation Veile et
al.\ observe by 72~h. This five-parameter model reduces the data
$\chi^2$ from 7.15 (M1) to 5.79, and does so while returning
$\mathrm{PBR}_\text{eff}$ to order unity, where M1 needed 2.43 to
compensate for the missing initial precipitate population. The six-parameter M5, which adds a
free dissolution rate on top of M4, reaches the same $\chi^2$ with that
rate indistinguishable from zero; it is carried forward only as the
alternative long-time branch it implies (finite saturation instead of
unbounded logarithmic growth, Section~\ref{sec:predictions}), not as a
competing fit. For M3 and M5 the free-parameter count equals the six
data points and the residual degrees of freedom are zero; their $\chi^2$
is nonetheless nonzero because the objective~\eqref{eq:objective} carries
one prior residual in addition to the six data residuals, so the fit is
not free to interpolate. No goodness-of-fit statistic is quoted for those
rows.

\begin{table}[htbp]
\centering
\caption{Layer-thickness data of Veile et al.~\citep{veile2024} (their
Fig.~9) used in this work: mean $\pm$ population standard deviation
over $n$ EDX line scans. The fits use standard errors
$\sigma = \mathrm{SD}/\sqrt{n}$. The Ni enrichment zone is not a
growing PDM layer and enters only the analysis of
Section~\ref{sec:nickel}.}
\label{tab:data}
\begin{tabular}{lcccc}
\toprule
Layer & $t$ (h) & mean (nm) & SD (nm) & $n$ \\
\midrule
Cr (barrier) & 72 & 37.3 & 7.7 & 3 \\
 & 168 & 98.3 & 21.7 & 4 \\
 & 480 & 134.3 & 5.4 & 3 \\
Fe (outer) & 72 & 143.3 & 111.5 & 3 \\
 & 168 & 225.5 & 142.7 & 4 \\
 & 480 & 240.3 & 172.5 & 3 \\
Ni (zone) & 72 & 31.3 & 3.4 & 3 \\
 & 168 & 43.0 & 11.1 & 4 \\
 & 480 & 36.0 & 5.0 & 3 \\
\bottomrule
\end{tabular}
\end{table}

\begin{table}[htbp]
\centering
\caption{Model ladder: nested apparent-parameter models fit to the
joint Cr+Fe layer-thickness data. The M3 row is the interior optimum
reached from the standard starting point, and it sits on the degenerate
ridge itself: the growth prefactor and dissolution rate diverge together
to $\sim\!10^{3}$~nm/h while the field parameter collapses to
$-4.4\times10^{-6}$~nm$^{-1}$. It attains the lowest data $\chi^2$ in
the ladder and the least defensible physics, predicting 155~nm at ten
years against M4's 535~nm; it is listed for completeness and rejected on
structure, not on fit.
The power law appears twice: evaluated at Veile et al.'s published
scan-level coefficients, and refit under the same weighted
objective~\eqref{eq:objective} as the mechanistic models. For M3 and
M5, $n_\text{free}$ equals the number of data points and dof $=0$.
$^\ddagger$The published-coefficient power law is \emph{evaluated}, not
fitted, against these six means (its coefficients were determined elsewhere, on individual scans in unweighted linear space), so it costs
no degrees of freedom here and its residual dof is 6. No information
criterion is computed for that row.}
\label{tab:ladder}
\begin{tabular}{lccccc}
\toprule
Model & Extra params & $n_\text{free}$ & $\chi^2_\text{data}$ & dof & $A_\text{bl}$ (nm/h) \\
\midrule
M1 core & --- & 4 & 7.15 & 2 & 0.7516 \\
M2 & $+C_\text{bl}$ & 5 & 7.15 & 1 & 0.7516 \\
M3 (degenerate) & $+C_\text{bl}, +C_x$ & 6 & 4.11 & 0 & 955.8 \\
\textbf{M4 (accepted)} & $+L_\text{ol,0}$ & 5 & \textbf{5.79} & 1 & \textbf{0.7269} \\
M5 & $+L_\text{ol,0}, +C_\text{bl}$ & 6 & 5.79 & 0 & 0.7269 \\
Power law (published coeffs) & $k,n$ per layer & 0$^\ddagger$ & 20.10 & 6$^\ddagger$ & --- \\
Power law (weighted refit) & $k,n$ per layer & 4 & 7.84 & 2 & --- \\
\bottomrule
\end{tabular}
\end{table}

The comparison with the empirical power law is reported both ways.
Evaluated at Veile et al.'s published per-layer coefficients the power
law gives $\chi^2 = 20.10$ against the joint layer means, but those
coefficients were fit to individual scans in unweighted linear space;
refit under the weighted objective~\eqref{eq:objective} it reaches
$\chi^2 = 7.84$ with four free parameters. On information criteria the
two descriptions are then statistically indistinguishable on this
dataset (Akaike criterion 15.8 for both; Bayesian criterion 14.8
mechanistic versus 15.0 power law; the small-sample-corrected criterion
is undefined for M4 at $n=6$, $n-k-1=0$, and is 55.8 for the power
law). These criteria are quoted as indicative only at this sample size.
The mechanistic model's advantage on the calibration window is not fit
quality (no comparison on three exposure times could demonstrate
that), but the physical coupling it enforces: the constant-volume
constraint ties the two layers through a single
growth-and-precipitation picture, where the power law treats each layer
as an independent curve (Fig.~\ref{fig:fits}); it is this coupling that
drives the divergent extrapolations of
Section~\ref{sec:predictions}.

On absolute goodness of fit, $\chi^2 = 5.79$ corresponds to $p = 0.016$
at one residual degree of freedom and $p \approx 0.12$ at three (the
effective-dof bracket of Section~\ref{sec:stats}). If the mild misfit
is real, the reported standard errors of the layer means understate the
scan-to-scan variability (consistent with the physical heterogeneity
Veile et al.\ report), and an error-scale correction of $\hat{s} = \sqrt{\chi^2/\text{dof}} = 1.4$--$2.4$ would widen every
uncertainty interval below by that factor. Unscaled intervals are
quoted throughout; no identifiability verdict changes under the maximal
scaling (the sharply identifiable parameters remain so at $2.4\times$
wider intervals, and the unidentifiable ones cannot get worse). The
factor is carried forward for one purpose: it sets how far the ten-year
band of Section~\ref{sec:predictions} can be inflated before the
comparison that section rests on would fail, and that comparison
survives the inflation.

The recovered kinetics place the apparent growth prefactor at
$A_\text{bl} = 0.727$~nm/h and the field exponent at
$b_3 = -0.0125$~nm$^{-1}$ ($-1.25\times10^{5}$~cm$^{-1}$). The
recovered $\mathrm{PBR}_\text{eff} = 1.096$ carries no prior and is
therefore a statement of the likelihood alone. It sits near unity, and
roughly a factor of two below the value a chromium mass balance on this
alloy implies for an FeCr$_2$O$_4$ barrier. That comparison does not
survive scrutiny in either direction, since the mass-balance figure is
phase-dependent, and the fitted parameter is not the same quantity once
outer-layer loss is admitted; it is taken up in
Section~\ref{sec:discussion}.

\subsubsection*{Where the fit statistic comes from}

\begin{figure}[htbp]
\centering
\includegraphics[width=0.86\textwidth]{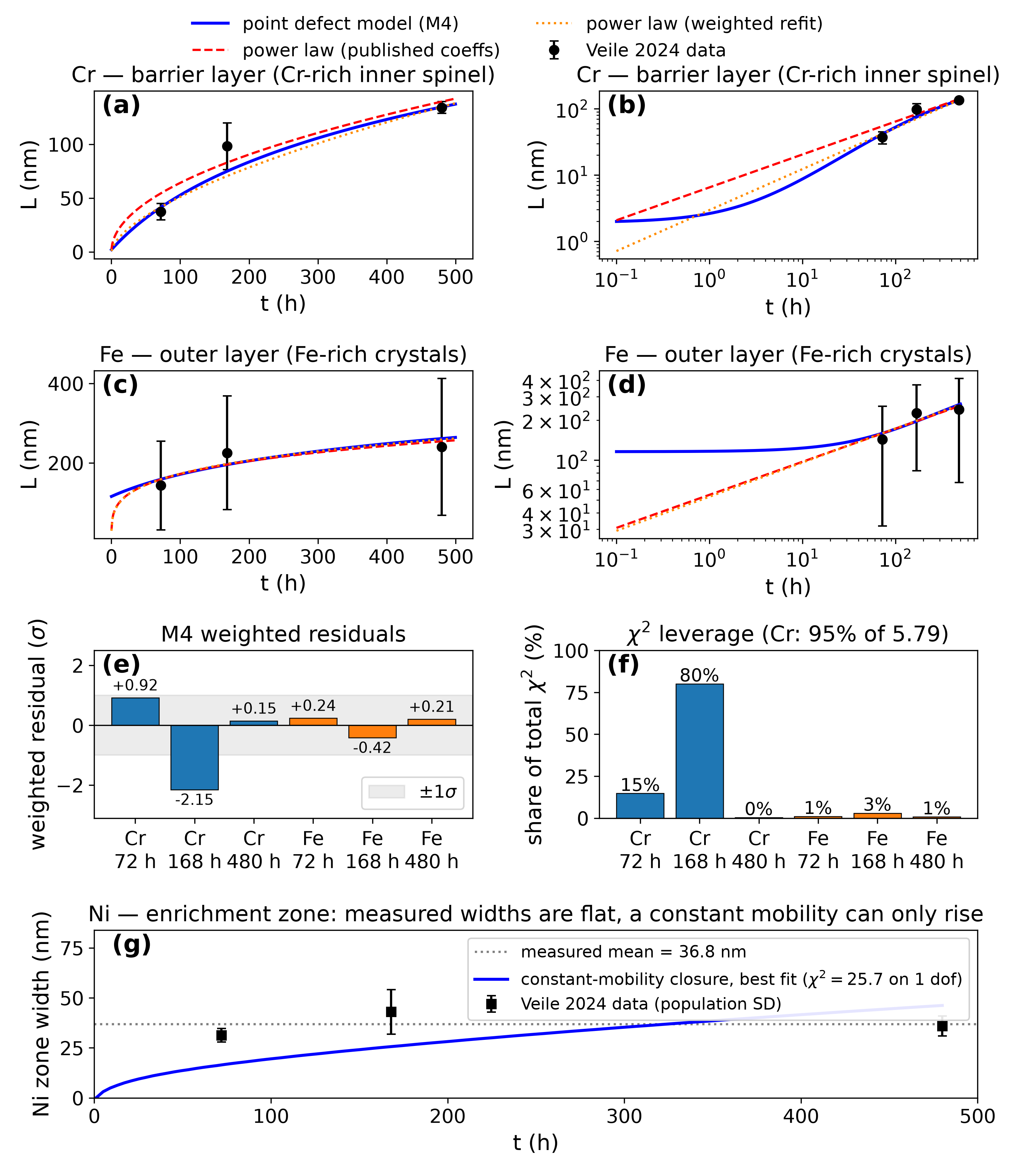}
\caption{M4 fit and the empirical power law against the measured
layer-thickness data. (a, b)~Cr barrier layer, linear and log--log
axes; (c, d)~Fe outer layer, linear and log--log axes. Points with
error bars in (a)--(d) are the layer means $\pm$ population standard
deviations of Table~\ref{tab:data}; the fits weight residuals by the
standard errors $\mathrm{SD}/\sqrt{n}$, so the plotted bars are wider
than the quantity actually minimized. Three curves are drawn in each of
(a)--(d): the blue solid curve is the M4 mechanistic fit, the red dashed
curve the power law evaluated at Veile et al.'s published coefficients,
and the orange dotted curve the power law refit under the weighted
objective~\eqref{eq:objective}.
(e)~Weighted residuals (model minus measurement, in units of the
standard error) of the M4 fit at the six data points, with the shaded
band marking $\pm1\sigma$, and (f)~each point's share of the total
$\chi^2$; bars are colored by layer in both, blue for chromium and
orange for iron. The chromium layer carries
95\% of the fit statistic and the 168~h chromium point alone 80\%, with
the $+$, $-$, $+$ chromium residual pattern indicating a systematic
curvature miss rather than scatter (Table~\ref{tab:leverage}). (g)~Ni enrichment-zone
width versus exposure time: black squares are the measured widths, the
dotted gray line their mean, and the blue curve the best-fit
constant-mobility transport closure of Section~\ref{sec:nickel}
evaluated densely in time. The rejection there is one of shape: the
closure can only rise, while the three measured widths are flat within
scatter across a factor of 6.7 in exposure time. As in Veile et al.'s
own analysis, this zone is not a growing PDM layer, and the fit is shown
to display its failure, not as an accepted description; error bars here
are population standard deviations, the weighting that fit uses.}
\label{fig:fits}
\end{figure}

A five-parameter model fitted to six means invites the question of which
data points actually carry the fit. Decomposing $\chi^2$ at the M4
optimum answers it (Fig.~\ref{fig:fits}e,f;
Table~\ref{tab:leverage}). The three chromium points contribute 95\% of
$\chi^2 = 5.79$ and the three iron points only 5\%: the outer layer's
scan-to-scan scatter is so large (standard errors of the mean of 32--45\%
of the mean, against 2--12\% for chromium) that it constrains the fit
only weakly. Within the chromium set the misfit is concentrated in a
single datum: the 168~h point, at $-2.15\sigma$, supplies 80\% of the
total. The residual signs run $+$, $-$, $+$ across the three
exposures, a systematic pattern rather than scatter, indicating that the
\emph{curvature} of the fitted logarithmic law does not track the three
measured chromium thicknesses.

Two consequences follow, and both are carried through the rest of the
paper. First, the curvature parameter $b_3$ is what governs the long-time
extrapolation of Section~\ref{sec:predictions}, so a systematic curvature
residual on the only three points that constrain it is the dominant
structural caveat on the ten-year number, and in our view larger than the
parametric uncertainty the bootstrap propagates. Second, the
constant-volume coupling and the initial-precipitate parameter
$L_\text{ol,0}$, although motivated by the iron layer, are in practice
constrained by it only weakly; the mechanistic reading of the iron
exponent in Section~\ref{sec:discussion} should be weighed accordingly.
Neither observation is grounds for preferring the power law, whose own
free parameters are fitted against the same six means with the same
leverage imbalance, but both bound how much any description calibrated on
this dataset can claim.

\begin{table}[htbp]
\centering
\caption{Residual and $\chi^2$ leverage decomposition of the accepted M4
fit. $\sigma$ is the standard error of the layer mean
(Table~\ref{tab:data}). The chromium layer carries 95\% of $\chi^2$ and
the 168~h chromium point alone carries 80\%.}
\label{tab:leverage}
\begin{tabular}{lcccrr}
\toprule
Layer & $t$ (h) & $\sigma$ (nm) & Model (nm) & Residual ($\sigma$) & \% of $\chi^2$ \\
\midrule
Cr (barrier) & 72 & 4.45 & 41.4 & $+0.92$ & 14.7 \\
 & 168 & 10.82 & 74.9 & $-2.15$ & 80.1 \\
 & 480 & 3.14 & 134.8 & $+0.15$ & 0.4 \\
Fe (outer) & 72 & 64.37 & 158.9 & $+0.24$ & 1.0 \\
 & 168 & 71.35 & 195.6 & $-0.42$ & 3.0 \\
 & 480 & 99.59 & 261.2 & $+0.21$ & 0.8 \\
\midrule
\multicolumn{4}{l}{Total $\chi^2 = 5.79$} & Cr 5.51 & 95.2 \\
\multicolumn{4}{l}{} & Fe 0.28 & 4.8 \\
\bottomrule
\end{tabular}
\end{table}

\subsection{Parameter identifiability and uncertainty quantification}\label{sec:identifiability}

\begin{figure}[htbp]
\centering
\includegraphics[width=\textwidth]{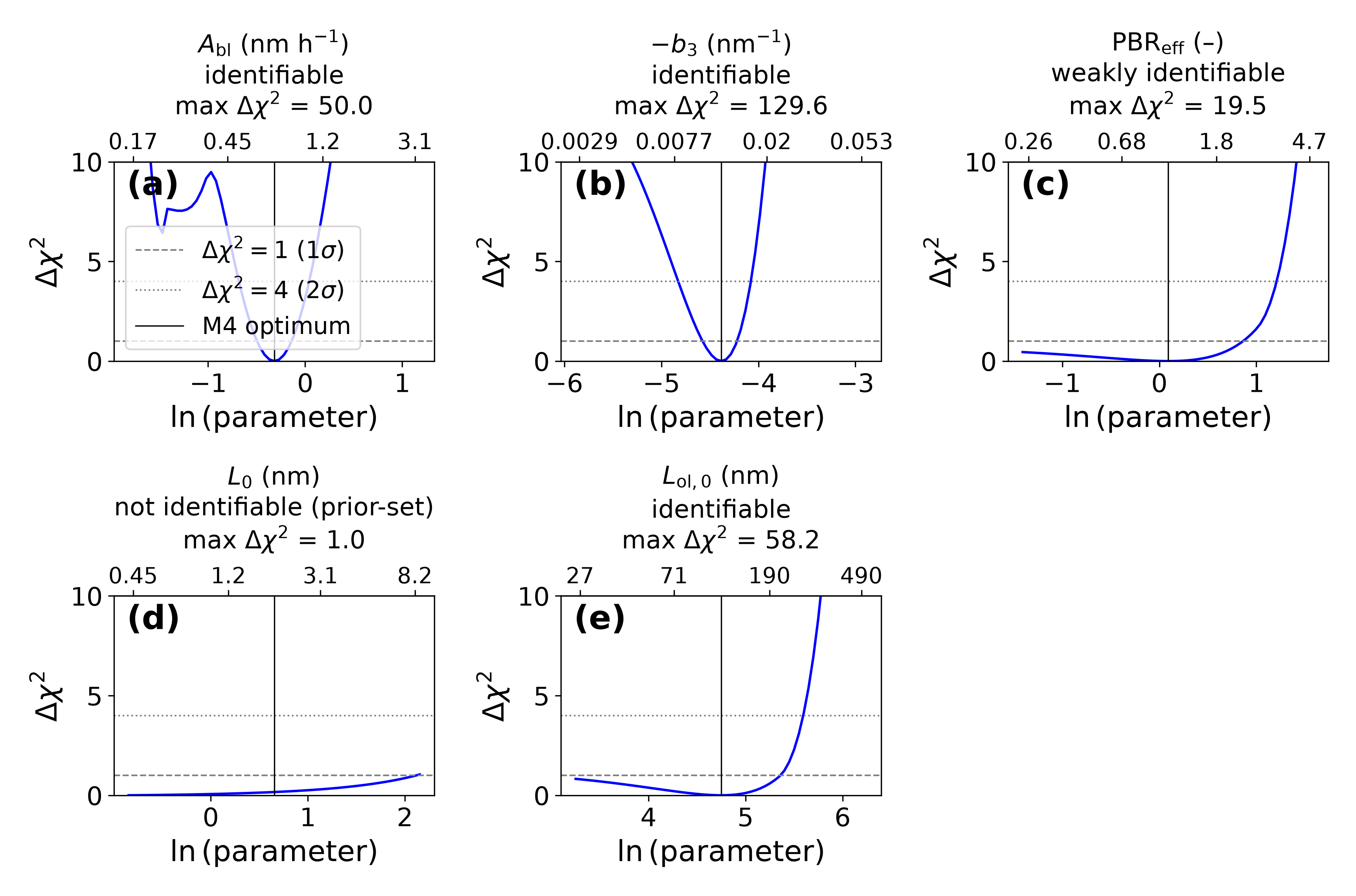}
\caption{Profile-likelihood identifiability map: $\Delta\chi^2$ over
the data residuals versus each M4 parameter, with the 1$\sigma$
($\Delta\chi^2 = 1$) and 2$\sigma$ ($\Delta\chi^2 = 4$) thresholds
marked. Panels are the five M4 parameters in the order of
Table~\ref{tab:params}: (a)~$A_\text{bl}$, (b)~$-b_3$ (the magnitude is
plotted, $b_3$ itself being negative throughout), (c)~$\mathrm{PBR}_\text{eff}$,
(d)~$L_0$, (e)~$L_\text{ol,0}$. In every panel the blue curve is the
profile, obtained by fixing the named parameter at each grid value and
re-optimizing all the others; the vertical solid line marks the M4
optimum, and the dashed and dotted horizontal lines the two thresholds.
The lower abscissa is the fitted coordinate, the natural log of
the parameter magnitude; the upper abscissa is the parameter itself, in
the units given in the panel title. Each panel title also carries the
resulting verdict and the maximum
$\Delta\chi^2$ attained over the scanned range: a profile that stays
flat is the signature of a parameter the data do not determine. The secondary structure
at low $A_\text{bl}$ in (a) is re-optimization of the remaining
parameters along a shallow direction; it lies far above both thresholds
and does not affect the interval.}
\label{fig:profile}
\end{figure}

If three chromium points carry the fit, the question that decides what
the fit is worth is which of the five parameters those points actually
determine, and the answer governs every extrapolation in
Section~\ref{sec:predictions}, since a parameter the data do not
constrain cannot be relied on outside the window that failed to
constrain it. Profile-likelihood scans referenced to the joint global
minimum (Fig.~\ref{fig:profile}, Table~\ref{tab:params}) show that the
three exposure times sharply constrain three of the five M4 parameters:
the growth prefactor, field exponent, and initial outer-layer thickness
(maximum $\Delta\chi^2$ of 50, 130, and 58 across their scans). The
Pilling--Bedworth ratio is only weakly identifiable (maximum
$\Delta\chi^2$ 19.5, interval spanning an order of magnitude), and the
initial barrier-layer thickness is formally unidentifiable from the
exposure-time data (maximum $\Delta\chi^2$ of 1.0; its quoted interval
simply reflects the $\pm1.5$ natural-log scan range: the parameter is
unbounded within the scanned range and is set by the air-film prior). A
shallow secondary structure in the growth-prefactor profile at low
parameter values is a local optimum of the compensating parameters; it
sits more than six $\Delta\chi^2$ units above the global minimum and
does not affect the quoted interval. Separately, an explicit search for
a dissolution-rate upper bound from the thickness data finds no bound
below 1~nm/h: the model's two qualitatively different long-time
behaviors remain numerically indistinguishable within the exposure
window, a point taken up in Section~\ref{sec:predictions}.

\begin{table}[htbp]
\centering
\caption{M4 parameters: fitted value, profile-likelihood 1$\sigma$
interval, 1000-member bootstrap mean $\pm$ standard deviation, and
identifiability verdict. The derived field strength is
$\varepsilon_f = 1.7\times10^4$~V/cm at the reference transfer
coefficient $\alpha_3 = 0.12$; its dominant (systematic) uncertainty is
the assumed $\alpha_3$, discussed in Section~\ref{sec:discussion}.
$^\dagger$For $L_0$ the bracket gives the scan-range endpoints, not
$\Delta\chi^2 = 1$ crossings: the profile never exceeds
$\Delta\chi^2 = 1$ anywhere in the scanned range. No parameter carries a
prior except $L_0$; $\mathrm{PBR}_\text{eff}$ is determined by the
likelihood alone.}
\label{tab:params}
\footnotesize
\setlength{\tabcolsep}{4pt}
\begin{tabular}{lccccl}
\toprule
Parameter & Unit & M4 fit & Profile 1$\sigma$ & Bootstrap & Verdict \\
\midrule
$A_\text{bl}$ & nm/h & 0.7269 & [0.626, 0.845] & $0.735 \pm 0.122$ & identifiable \\
$b_3$ & nm$^{-1}$ & $-0.01249$ & [$-0.0145$, $-0.0108$] & $-0.01243 \pm 0.00202$ & identifiable \\
$\mathrm{PBR}_\text{eff}$ & --- & 1.096 & [0.245, 2.32] & $1.169 \pm 0.931$ & weakly identifiable \\
$L_0$ & nm & 1.9237 & [0.429, 8.20]$^\dagger$ & $1.923 \pm 0.0358$ & not identifiable (prior-set) \\
$L_\text{ol,0}$ & nm & 115.58 & [25.79, 210.6] & $110.5 \pm 74.7$ & identifiable \\
\bottomrule
\end{tabular}
\end{table}

For a parameter carrying a prior, the profile alone cannot separate data
leverage from prior leverage, because both are centered where the prior
is. The direct test is to move the prior and see whether the estimate
follows (Table~\ref{tab:priortest}). $L_0$ follows it completely:
widening its prior fivefold moves the estimate from 1.92 to 1.14~nm
(41\%), and removing it drives $L_0$ to zero, the likelihood having no
interior optimum at all. $\mathrm{PBR}_\text{eff}$ behaves in the
opposite way. It carries no prior here, so the test is run in reverse:
imposing the prior this analysis previously used, centered on 1.05 with
$\sigma_{\ln} = 0.20$, moves the estimate only from 1.096 to 1.051
(4.1\%) and leaves the data $\chi^2$ at 5.79. The likelihood therefore
does define an interior optimum for this parameter, and the former prior
was displacing it by about 4\% rather than creating it. This is the test
that licenses reading $\mathrm{PBR}_\text{eff} \approx 1.10$ as a weak
but genuine data statement while reading $L_0$ as prior-set.

\begin{table}[htbp]
\centering
\caption{Prior-leverage test. Each row refits M4 with the Gaussian
log-space prior widths of Eq.~\eqref{eq:objective} modified as shown
(``off'' = prior removed). A data-determined parameter does not move when
its prior is relaxed; a prior-set one drifts or loses its optimum.}
\label{tab:priortest}
\begin{tabular}{lccccc}
\toprule
Prior setting & $\sigma_{\ln \mathrm{PBR}}$ & $\sigma_{\ln L_0}$ & $\mathrm{PBR}_\text{eff}$ & $L_0$ (nm) & $\chi^2_\text{data}$ \\
\midrule
As fitted (baseline) & --- & 0.60 & 1.096 & 1.924 & 5.79 \\
Impose PBR 1.05, $\sigma=0.20$ & 0.20 & 0.60 & 1.051 & 1.924 & 5.79 \\
Impose PBR 1.05, $\sigma=1.00$ & 1.00 & 0.60 & 1.069 & 1.924 & 5.79 \\
$L_0$ prior $\times5$ & --- & 3.00 & 1.097 & 1.144 & 5.71 \\
$L_0$ prior off & --- & --- & 1.098 & 0.000 & 5.58 \\
Both (impose PBR, $L_0$ off) & 0.20 & --- & 1.051 & 0.000 & 5.59 \\
\bottomrule
\end{tabular}
\end{table}

\begin{figure}[htbp]
\centering
\includegraphics[width=\textwidth]{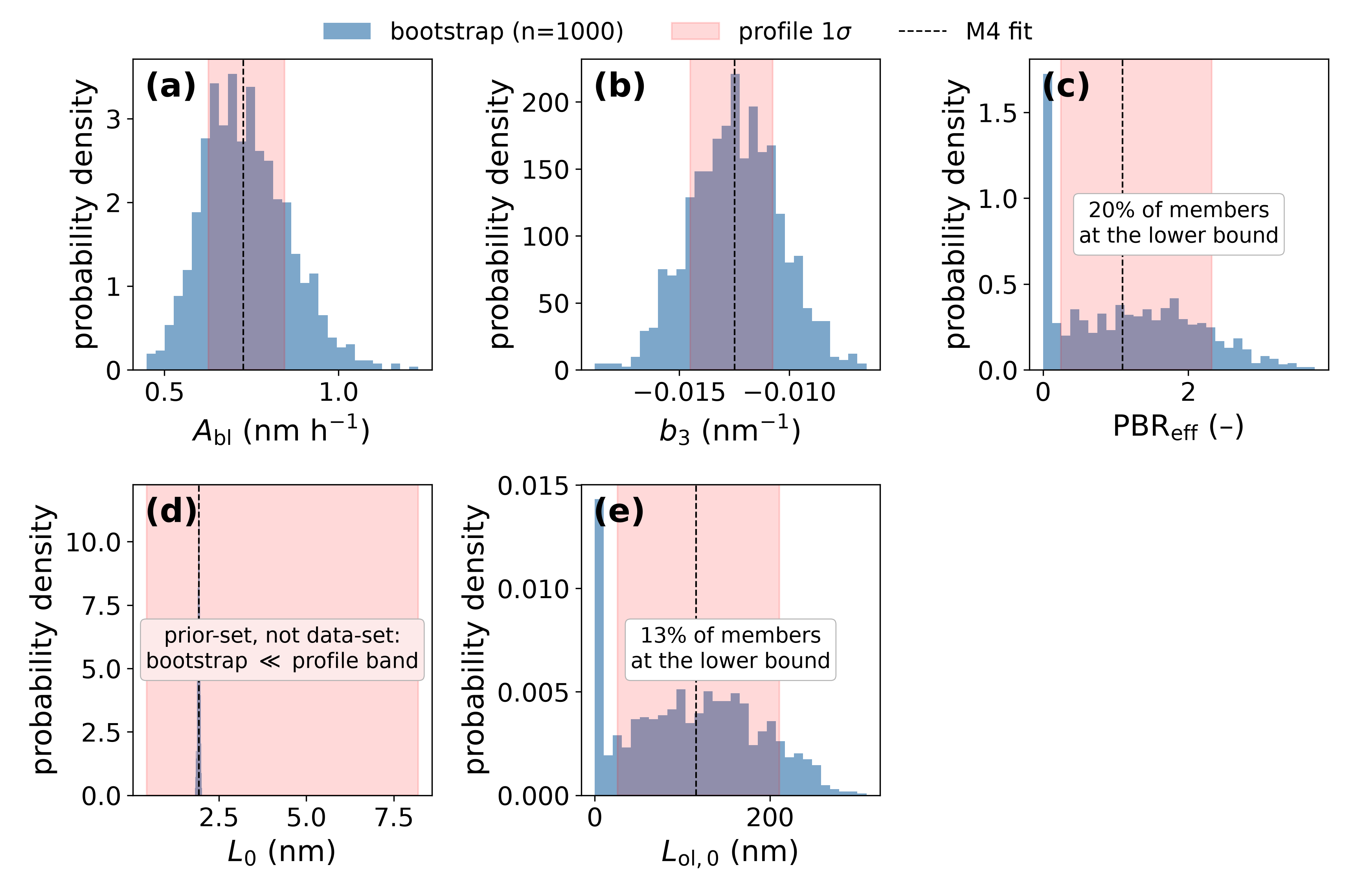}
\caption{Bootstrap ensemble versus profile likelihood: 1000-member
parametric bootstrap distribution for each M4 parameter, compared
against the profile-likelihood 1$\sigma$ band. Panels follow the order
of Table~\ref{tab:params}: (a)~$A_\text{bl}$, (b)~$b_3$,
(c)~$\mathrm{PBR}_\text{eff}$, (d)~$L_0$, (e)~$L_\text{ol,0}$. In each,
the blue histogram is the distribution of the 1000 refitted values
(plotted as a probability density, so panel heights are not comparable
across parameters), the red shaded band is the profile-likelihood
1$\sigma$ interval of Table~\ref{tab:params}, and the dashed vertical
line is the M4 point estimate. Two annotated features
are results rather than plotting artifacts. The spike at the lower bound
in (c) and (e) is the fraction of resampled datasets for which the
parameter is driven to zero, which is what weak identifiability looks
like in a resampling ensemble. In (d) the bootstrap is far narrower than
the profile band: resampling the data cannot explore a direction the
data do not constrain, so the narrow histogram is not a tighter
constraint but the signature of a prior-set parameter.}
\label{fig:ensemble}
\end{figure}

The 1000-member parametric bootstrap cross-validates these conclusions
by an independent method Fig.~\ref{fig:ensemble}, and the two weakly
constrained parameters fail it in opposite directions. For the sharply
identifiable growth prefactor and field exponent the bootstrap and
profile intervals agree closely. For $L_0$ the bootstrap spread
($\pm0.036$~nm) is far narrower than the profile interval; this is not a
tighter constraint but an artifact of the bootstrap perturbing only the
data while the profile explores the full parameter range: when the data
have no leverage, every resampled refit collapses onto the same prior.
$\mathrm{PBR}_\text{eff}$, now unconstrained by any prior, shows the
opposite signature: a bootstrap standard deviation of 0.93 on a central
value of 1.10, with the estimate collapsing toward zero in a sizeable
minority of replicates (16th percentile $5\times10^{-9}$). Its profile
interval [0.245, 2.32] and its bootstrap spread agree that three
exposure times pin this parameter only to within about an order of
magnitude. The initial outer-layer thickness sits between these cases:
its profile is sharp ($\Delta\chi^2 = 58$) but its bootstrap is not
($110 \pm 75$~nm, 16th percentile 17~nm), because $L_\text{ol,0}$ and
$\mathrm{PBR}_\text{eff}$ trade off against each other in the
outer-layer data and a resampled replicate can move along that trade-off
at little cost in $\chi^2$. The profile-likelihood interval, not the
bootstrap spread, is the honest uncertainty statement for $L_0$; for
$\mathrm{PBR}_\text{eff}$ the two methods agree that the constraint is
weak; and for $L_\text{ol,0}$ the wider bootstrap should be preferred,
since it is the one that samples the trade-off. All three verdicts are
consequences of the same shortage of exposure times.

\subsection{Long-time extrapolation and operating-envelope sensitivity}\label{sec:predictions}

The identified kinetics are wanted for a horizon no exposure in this
dataset reaches, and it is there that the mechanistic and empirical
descriptions part company. The result of this section is that separation
rather than the thickness it brackets: the point estimate inherits every
identifiability limitation cataloged above, while the separation
survives them.

\begin{figure}[htbp]
\centering
\includegraphics[width=\textwidth]{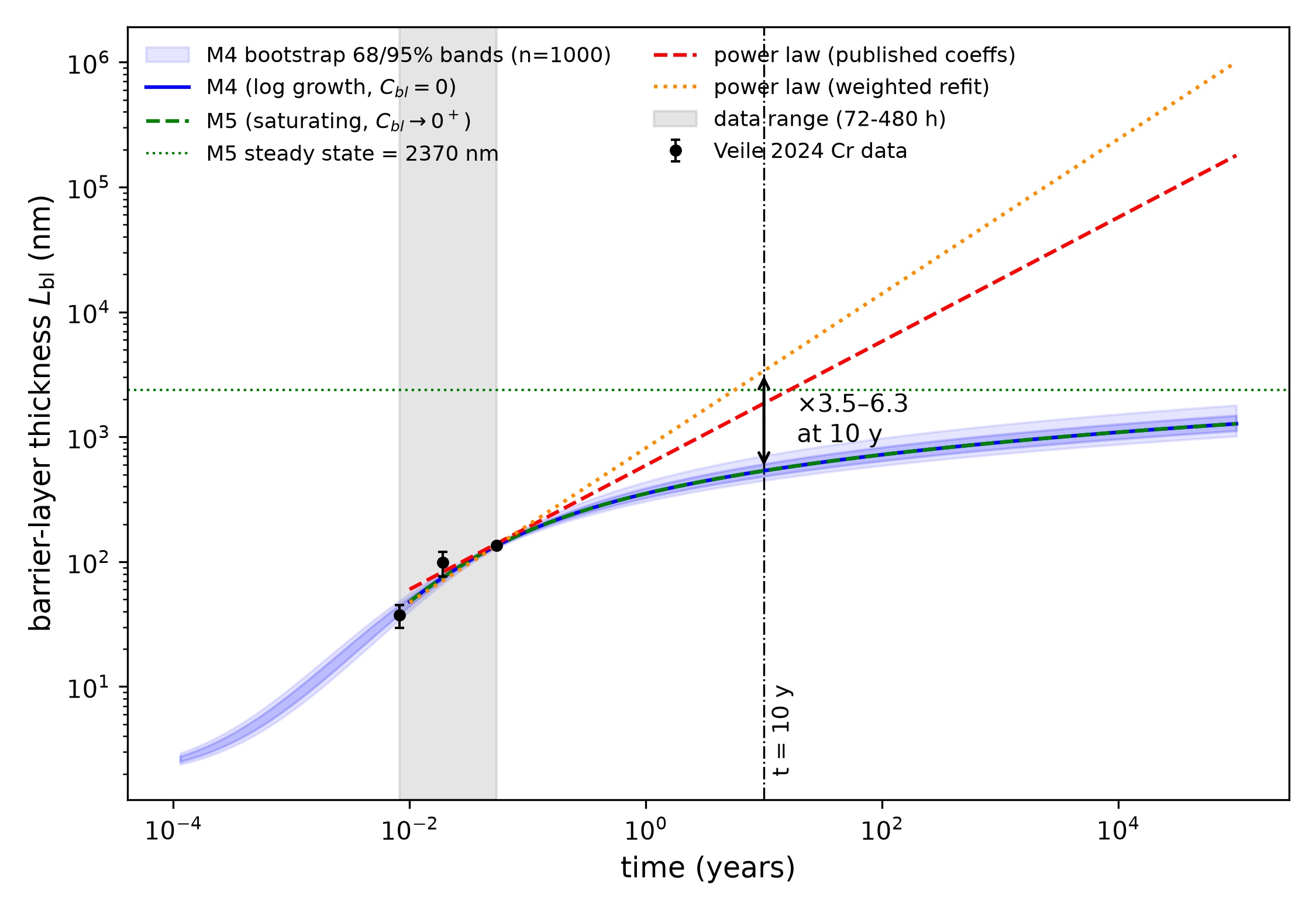}
\caption{Long-time prediction with propagated parametric uncertainty:
M4 (blue solid, logarithmic growth) versus M5 (green dashed, saturating
branch, with its steady-state thickness marked by the green dotted
horizontal line) versus the two power-law variants (red dashed at the
published coefficients, orange dotted for the weighted refit), to
$10^5$~years on log--log axes. Shaded bands are the 68\% and 95\%
intervals of the 1000-member parametric bootstrap propagated through
the M4 closed form, conditional on the M4 model structure and the
reported standard errors; the M5 curve is indistinguishable from M4 at
all feasible exposures. The three digitized chromium measurements are
plotted as black circles and the 72--480~h calibration window is shaded
gray, so the interval the extrapolations leave is visible rather than
implied. The vertical line marks the ten-year horizon,
and the arrow there is the separation the paper's conclusion rests on:
the two power-law variants lie a factor of 3.5 to 6.3 above the
mechanistic barrier layer and outside its 95\% band.}
\label{fig:longtime}
\end{figure}

At ten years (a factor of roughly 180 beyond the 480-h calibration
window), the identified model predicts a barrier-layer thickness of
535~nm, with bootstrap 68\% and 95\% intervals of [482, 606] and [447,
704]~nm (Fig.~\ref{fig:longtime}). These bands quantify parametric
uncertainty conditional on the M4 model structure and on the reported
standard errors; under the maximal error-scale correction of
Section~\ref{sec:ladder} ($s = 2.4$, applied in log space) the 95\%
interval widens to approximately [340, 1040]~nm. Extrapolating Veile
et al.'s published power-law coefficients over the same horizon instead
predicts 1853~nm (a factor of 3.5), and the weighted-refit power law
3375~nm (a factor of 6.3; bootstrap 68\% interval on the ratio [5.6,
7.0]). Either power-law prediction lies outside the mechanistic band
even after the maximal inflation. The divergence is the practical
consequence of the constant-volume-coupled growth law saturating
logarithmically while the power law, fit independently to each layer,
extrapolates indefinitely.

The model's own two long-time branches, unbounded logarithmic growth
(M4) versus eventual saturation (M5), track within about five percent
of each other out to $10^5$~years, which is well inside the 2--12\% standard error of the chromium layer means: no exposure
experiment at the precision this technique delivers could distinguish
them from thickness data, the prediction-space
restatement of the dissolution-rate unidentifiability of
Section~\ref{sec:identifiability}. Discriminating mechanistic from
power-law extrapolation is, by contrast, feasible. Under a conservative
scatter model for a future campaign (the worst observed chromium
scan-level relative standard deviation, 22\%, averaged over three
scans, giving a 12.7\% standard error), the separation between the two
predictions first exceeds twice that scatter at 2546~h with the
published coefficients (1437~h refit) and three times at 4014~h
(2020~h refit). These thresholds compare the separation to future
measurement scatter alone, and so understate the required exposure. Adding
the mechanistic prediction's own parametric uncertainty (the
1000-member bootstrap propagated to each exposure time) and, for the
refit, the power-law fit uncertainty (an equivalent bootstrap of the same
weighted objective) in quadrature moves the 2$\sigma$ threshold to 2988~h
with the published coefficients and 1928~h with the refit, and the
3$\sigma$ threshold to 5129~h and 5493~h respectively. No fit uncertainty
is carried on the published-coefficient branch, since those coefficients
were determined elsewhere and no covariance for them is reported; that
branch is correspondingly optimistic at 3$\sigma$ and the refit branch is
the one to plan against.

The 3$\sigma$ figures carry a qualification the 2$\sigma$ figures do not,
and it is the same lesson the rest of the paper reaches by other routes.
Once the power law's own extrapolation uncertainty is carried, that
uncertainty grows in proportion to the prediction it attaches to, so the
separation measured in standard deviations is not monotone in exposure
time. Against the refit it crosses 3$\sigma$ at 5493~h, peaks at
3.18$\sigma$ near 12000~h, and falls back through 3$\sigma$ at
34400~h to 2.89$\sigma$ by 50000~h. Three-sigma discrimination against
the refit is therefore available in a window of roughly 5500 to
34000~h and never afterwards, no matter how long the exposure runs; the
2$\sigma$ criterion, by contrast, once met is never lost. A designed
exposure of roughly 3000~h is the realistic requirement for 2$\sigma$
discrimination against the full error budget, and 3$\sigma$ requires
5000--6000~h and must be placed inside that window rather than simply
extended toward it. The scatter-only numbers should be read as the
optimistic bound they are.

The separation above is established at the one operating condition the
data were collected at, which leaves open how much of it survives a
change of condition. Because the dataset was collected at a single
temperature and oxygen level, the model's sensitivity is mapped across a
scanned operating envelope of 200--285$^\circ$C at 7~MPa and 0.01--8~ppm
dissolved oxygen, deliberately extended beyond normal-water-chemistry
oxygen levels ($\sim$0.2--0.4~ppm) to bound off-normal chemistry. The
upper temperature bound is the liquid-phase stability limit at 7~MPa
($T_\text{sat} \approx 286^\circ$C), so nominal BWR coolant temperature
($\approx 288^\circ$C at operating pressure) sits essentially at the
edge of the scanned box, not comfortably inside it. The temperature
channel follows the rate-constant structure of the supercritical-water
formulation~\citep{li2020scwpdm}: the rate-determining constant scales
as $\exp[-\alpha_3 \Delta G^0_R/R\,(1/T - 1/T_0)]$, where $\Delta G^0_R$
is the standard Gibbs energy of the growth reaction, $R$ the gas
constant, $T$ the absolute temperature and $T_0 = 513$~K the calibration
temperature, so that the effective Arrhenius activation energy is
$\alpha_3 \Delta G^0_R$; with the standard Gibbs energy of the growth
reaction unmeasurable from single-temperature data, $\Delta G^0_R$ must
be scanned explicitly. The scan is set by the effective activation
energy it implies rather than by $\Delta G^0_R$ itself: at $\alpha_3 =
0.12$ the range 25--417~kJ/mol corresponds to $\alpha_3 \Delta G^0_R =
3$--50~kJ/mol, which spans both the low values implied by transferring a
supercritical-water transfer coefficient and the several tens of kJ/mol
reported for weight-gain kinetics of austenitic steels in
high-temperature water. Scanning only the lower part of that range would
build the robustness conclusion on the assumption most favorable to it.
A field-attenuation channel is tied to the same temperature dependence,
and the dissolved-oxygen channel is an ideal-Nernst shift of the growth
prefactor, which is an explicit lower bound on the true
corrosion-potential response (Section~\ref{sec:discussion}).

\begin{figure}[htbp]
\centering
\includegraphics[width=\textwidth]{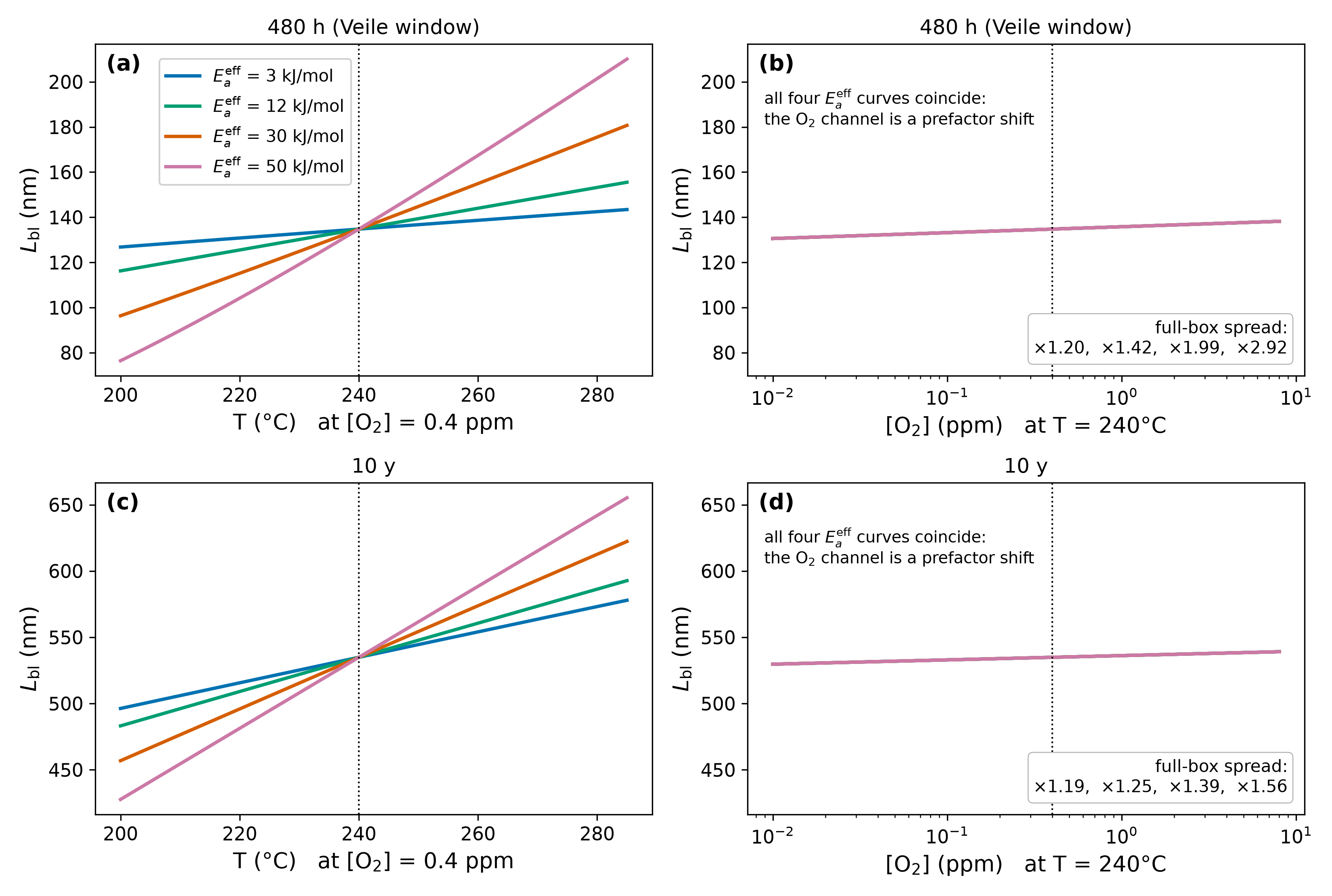}
\caption{Operating-envelope sensitivity of the predicted barrier-layer
thickness, at (a,~b)~480~h and (c,~d)~ten years. Left panels sweep
temperature at the calibration oxygen level; right panels sweep
dissolved oxygen across three decades at the calibration temperature,
on the same vertical scale. Each curve is one scanned value of the
effective activation energy $\alpha_3\Delta G^0_R$, spanning
3--50~kJ/mol; the dotted line marks the Veile et al.\ calibration
condition, through which every curve passes by construction. Temperature
carries the sensitivity and it grows with activation energy, while three decades of dissolved oxygen shift the prediction by under six percent at 480~h and under two percent at ten years, and do so identically at every activation energy, the oxygen channel being a prefactor shift. Boxed factors are the full-envelope spreads of the four plotted activation energies, from Table~\ref{tab:envelope}. Contour maps over the full
$(T, [\mathrm{O}_2])$ box are given in the SI (Fig.~S3).}
\label{fig:sensitivity}
\end{figure}

Across this envelope (Table~\ref{tab:envelope}; line sweeps in the Fig.~\ref{fig:sensitivity}, the predicted barrier-layer thickness varies
by a factor of 1.20 to 2.92 at 480~h and 1.19 to 1.56 at ten years,
depending on the scanned effective activation energy; temperature
dominates through the combined Arrhenius and field-attenuation channels,
and its sensitivity grows with activation energy, while three decades of
dissolved oxygen shift the prediction by under six percent at 480~h and
under two percent at ten years, and do so identically at every
activation energy, the oxygen channel being a prefactor shift. Two features of this
result matter more than the headline range. First, the sensitivity is
strongly conditional on the activation energy: over the low 3--12~kJ/mol
sub-range the ten-year envelope factor is a nearly negligible
1.19--1.25, but at 50~kJ/mol it is 1.56, and the 480-h factor roughly
doubles from 1.42 to 2.92. Second, the ten-year factor is consistently
smaller than the 480-h factor, because logarithmic growth compresses
prefactor differences at long times: the robustness of the
\emph{long-time} prediction is a property of the growth law, and is not
inherited from a robustness of the short-time behavior. The long-time
prediction is therefore moderately, not strongly, sensitive to operating
condition across the full plausible activation-energy range, and a
factor of about 1.6 should be carried alongside the parametric band of
Fig.~\ref{fig:longtime} rather than the 1.25 the narrow scan alone would
suggest.

\begin{table}[htbp]
\centering
\caption{Operating-envelope spread of the predicted barrier-layer
thickness over the scanned box (200--285$^\circ$C at 7~MPa, 0.01--8~ppm
O$_2$), as a function of the assumed effective Arrhenius activation
energy $\alpha_3\Delta G^0_R$. The first three rows are the
originally scanned range; the last two extend it to the values reported
for weight-gain kinetics on austenitic steels in high-temperature water.}
\label{tab:envelope}
\begin{tabular}{cccccc}
\toprule
$\Delta G^0_R$ & $\alpha_3\Delta G^0_R$ & \multicolumn{2}{c}{$L_\text{bl}$(480~h)} & \multicolumn{2}{c}{$L_\text{bl}$(10~y)} \\
\cmidrule(lr){3-4}\cmidrule(lr){5-6}
(kJ/mol) & (kJ/mol) & range (nm) & factor & range (nm) & factor \\
\midrule
25 & 3 & 123--147 & 1.20 & 491--583 & 1.19 \\
50 & 6 & 119--151 & 1.27 & 487--587 & 1.21 \\
100 & 12 & 112--159 & 1.42 & 478--597 & 1.25 \\
250 & 30 & 93--185 & 1.99 & 452--627 & 1.39 \\
417 & 50 & 73--214 & 2.92 & 423--660 & 1.56 \\
\bottomrule
\end{tabular}
\end{table}

\subsection{Spatial extension: defect transport and composition}\label{sec:spatial}

Veile et al.'s line scans resolve full composition-versus-depth
profiles that the zero-dimensional fit never uses. We therefore extend
the model spatially (resolving oxygen- and cation-vacancy transport
across the barrier layer under the already-identified kinetics), and
use the resulting composition predictions as an independent cross-check
against data the fit never saw. The transport problem admits no
differentiable-integrator shortcut of the kind the zero-dimensional
kinetics allow, so the spatial solves use the soft-mode field
representation; it is verified against an independent Newton solve on
the same node set, which for this steady problem is both cheaper and
more accurate, so the comparison below grades the neural solver rather
than relying on it. Every kinetic parameter entering this section is
frozen at its zero-dimensional value.

That cross-check is only worth running on quantities the measurement
could resolve, so before any spatial inversion a synthetic
identifiability gate asks which spatial quantities the dataset could
constrain at all, using an
instrument model with position jitter, compositional noise, and,
critically, the scan-to-scan physical heterogeneity Veile et al.'s
replicate scans exhibit (SI, Section~S2.5; omitting the heterogeneity
term produces a spuriously optimistic verdict for every parameter).
Twenty replicate synthetic experiments (SI, Table~S7) show that
the barrier-layer thickness, interfacial transition widths, nickel-zone
amplitude and width, and in-layer chromium fraction are identifiable
(replicate relative standard deviations 2--19\%), while the in-layer
defect-transport composition gradient, the one quantity the spatial
extension uniquely adds, is weak (47\%, with a minimum statistically
detectable gradient of 0.93 absolute weight percent across the layer).
The spatial model's inverse scope is restricted accordingly, and the
defect-transport gradient is treated as a forward prediction, not a
fitting target.

The steady-state transport problem, solved by Newton iteration on a
spectral collocation grid and by an NSEM forward solve, agrees with its
analytic reference to $8.9\times10^{-16}$ (oxygen vacancies) and
$2.2\times10^{-13}$ (cation vacancies); the NSEM solve agrees with the
Newton reference to $2.5\times10^{-4}$ and $5.4\times10^{-5}$ relative
error (SI, Fig.~S4). The verification also establishes a
structural identifiability result: the steady nondimensional profile
shape depends only on the P\'eclet-type group
$\mathrm{Pe} = |z|\gamma \varepsilon_f L$ (in which $z$ is the charge
number of the migrating defect, $L$ the barrier-layer thickness, and
$\gamma$ and $\varepsilon_f$ are as above), which the zero-dimensional
fit already fixes, and not on the defect diffusivity. This was confirmed numerically,
not only by inspection of the equations: solving the identical problem
with the diffusivity varied over four orders of magnitude leaves the
profile shape unchanged to machine precision, while the absolute
concentration scale, invisible to a composition measurement such as
EDX, scales inversely with diffusivity exactly as the equations
predict. Quasi-steady composition measurements therefore cannot
constrain the defect diffusivity; only genuinely transient early-time
composition data would carry diffusivity sensitivity, through the
migration transit times quantified below. This structural argument
independently corroborates the statistical gate above.

Everything above treats the defect profile as quasi-steady, which is an
approximation every zero-dimensional PDM application makes implicitly
and none, as far as we are aware, has quantified: the profile is assumed
to equilibrate instantaneously to its steady shape as the layer grows.
The transient extension, solved on a moving Landau-transformed domain
tracking the growing layer, measures the error that assumption incurs.
At the end of the 480-h window the fast oxygen-vacancy species
(migration transit time of order 5~h) lags its quasi-steady shape by
0.69\%, and the slower cation-vacancy species (transit time of order
36~h) by 5.79\% (SI, Fig.~S2), both errors scaling with the ratio of
transit time to growth timescale as the physics predicts. Both are
inside the acceptance tolerances used here, so the quasi-steady
treatment of the composition prediction below is licensed rather than
assumed.

\begin{figure}[htbp]
\centering
\includegraphics[width=\textwidth]{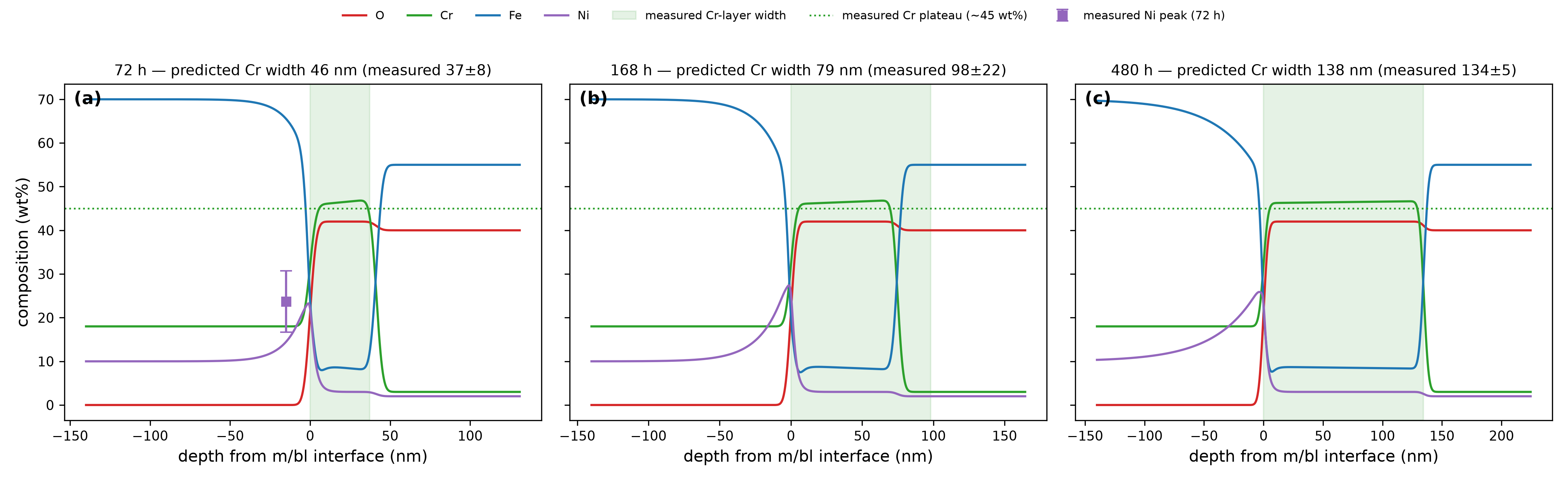}
\caption{Zero-free-parameter composition-map prediction against Veile
et al.'s measured EDX features at (a)~72, (b)~168, and (c)~480~h.
Curves are predicted composition profiles (O, Cr, Fe, Ni; weight
percent) versus depth from the metal/barrier-layer interface; overlays
mark the measured features compared: Cr-layer width (shaded band), Cr
plateau level ($\sim$45~wt\%, dotted), and the 72-h Ni-peak amplitude
($23.7\pm7.0$~wt\%, error bar). Full digitized EDX traces are not
available, so the comparison is feature-based.}
\label{fig:composition}
\end{figure}

The composition-map prediction uses no free parameters, and its two
components test different things. The predicted chromium plateau inside
the barrier layer, 46.7 weight percent against a measured plateau of
approximately 45 weight percent, follows from spinel stoichiometry alone, using the FeCr$_2$O$_4$ assignment~\citep{ziemniak2003spinel} Veile et
al.\ report as one of several candidate chromium-rich spinels (alongside
Cr$_2$O$_3$, Fe$_2$CrO$_4$, and NiCr$_2$O$_4$) consistent with their
X-ray photoelectron spectroscopy results, and therefore validates a
plausible phase assignment, not the kinetics. The kinetic content is in
the layer widths, which the frozen kinetics match to within $+22\%$,
$-20\%$, and $+3\%$ at the three exposure times
(Fig.~\ref{fig:composition}), agreement at the tens-of-percent level
from a zero-free-parameter prediction, not a quantitative
reproduction, with widths extracted from prediction and measurement by the same
criterion Veile et al.\ use (chromium trace exceeding its far-field
baseline by three weight percent). The predicted in-layer defect
gradient (1.45, 0.96, and 0.46 absolute weight percent at 72, 168, and
480~h) straddles the 0.93 weight percent detection floor from the
identifiability gate: marginally detectable at the earliest exposure,
undetectable by the latest.

\subsection{The nickel enrichment zone}\label{sec:nickel}

The nickel zone is the one feature of the duplex scale that the
identified kinetics do not describe, and the reason is instructive. An
exact moving-frame nickel-transport model with a constant effective
mobility (SI, Section~S3), fitted to the measured zone widths and the
72-h peak amplitude, is \emph{rejected} by the data: $\chi^2 = 25.7$ on
one degree of freedom. The rejection is one of shape, not scale: the
model predicts a zone width growing monotonically with exposure, while
the measured widths are approximately flat (31, 43, 36~nm at 72, 168 and
480~h; Fig.~\ref{fig:fits}g). Because parameter values and nested-model $\Delta\chi^2$
comparisons obtained under a structurally rejected model are not
interpretable as estimates or as evidence, the fitted values and the
nested-model comparison are reported in the SI (Section~S3.1,
Table~S3) and are not carried into the Discussion as findings.

Two statements about the nickel zone do survive independently of that
fit, and they are the section's result. First, the measured widths are
flat within scatter across a factor of 6.7 in exposure time, which no
constant-mobility transport model of this class can produce: any such
model predicts a width growing as the square root of time or faster.
Second, any effective mobility consistent with the observed zone
scale (a zone tens of nanometers wide developing over hundreds of
hours) is of order $10^{-17}$~cm$^2$/s, which exceeds extrapolated
lattice self-diffusion of nickel in austenite at 513~K by many orders of
magnitude. This is a scale argument, independent of the fitted value: the
zone-forming transport must be short-circuit or defect-flux mediated
rather than lattice-diffusive. Taken together, the two point to a
time-decaying or interface-coupled mobility.

That mechanism is exactly what three exposure times cannot resolve. Two
physically motivated one-parameter extensions were tested, a
defect-flux-coupled effective mobility tied to the identified growth
kinetics and a finite-capacity interfacial trapping closure, and both
fail a twenty-replicate synthetic identifiability gate before touching
real data (SI, Table~S4): the flux-coupled closure's new parameter is
recoverable but destabilizes the already-identified transport parameters
past threshold, and the finite-capacity closure's parameter is not
recoverable at all (SI, Section~S3.2). A forward-only scan across the
flux-coupled closure's plausible range confirms no single value is
simultaneously shape-correct
and a quantitative improvement. Resolving the nickel zone is therefore
left as a target for denser exposure-time data, and is one more instance
of the exposure-time limitation that recurs throughout this paper.

\subsection{Grading the inversion framework, and a failure mode specific
to sparse data}\label{sec:pinn}

Every verdict above was read off a single converged optimum, which
raises the question of how far that optimum depends on the machinery
that produced it. Because the reduced kinetics have a closed-form
solution, this problem
admits something that most physics-informed inverse problems do not: an
exact reference against which the neural inversion can be graded rather
than merely trusted. This section uses that reference twice: once to
establish that the hard-mode inversion is interchangeable with a
classical fitter, and once to expose a soft-mode failure that the
training loss does not reveal.

Applied to the real dataset, the hard-mode neural inversion recovers the
deterministic M4 optimum to within 0.6\% on every parameter, and its
trajectory $\chi^2$ of 5.79 matches the classical baseline exactly
(SI, Table~S6). Its recovery behavior was first validated on
synthetic data generated from known ground-truth kinetics with 2\%
relative noise at the same exposure times: recovery deviations are
1.37\% (growth prefactor), 3.39\% (field exponent), 0.59\%
(Pilling--Bedworth ratio), 11.06\% (initial barrier-layer thickness),
and 0.16\% (initial outer-layer thickness). The one large deviation
lands on exactly the parameter the profile-likelihood analysis flags as
unidentifiable, a direct cross-validation of the identifiability
finding by an independent method. A 50-member subset of the bootstrap
re-solved through the hard-mode inversion gives spreads statistically
consistent with the 1000-member deterministic ensemble (growth
prefactor $0.74 \pm 0.12$ deterministic, $0.76 \pm 0.14$
physics-informed), confirming the two fitters are interchangeable. Both neural backbones
also pass the 0.5\% forward acceptance criterion under identical
settings (relative $L^\infty$ errors 0.14\% and 0.12\% for the two
thickness fields with the multilayer-perceptron backbone, 0.11\% and
0.04\% with the Kolmogorov--Arnold-network backbone), a parity result:
on this problem, backbone choice does not materially change forward
accuracy.

\begin{figure}[htbp]
\centering
\includegraphics[width=\textwidth]{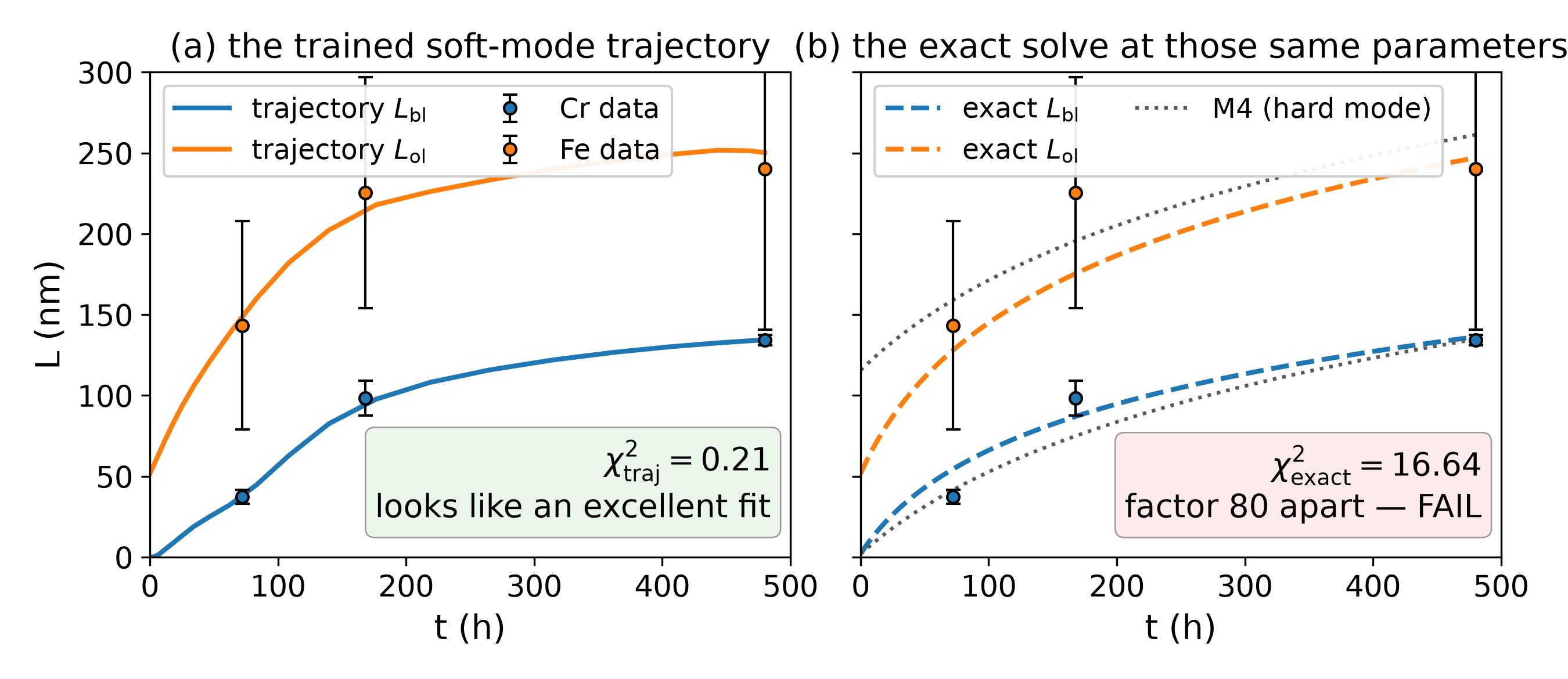}
\caption{The F-1 failure mode. Both panels use the same six measurements
and the same recovered parameters; in both, blue denotes the barrier
layer and orange the outer layer, and the points are the measured means
with their standard errors. (a)~The trained soft-mode trajectory (solid
curves), which is what the data loss sees: it passes through every
point, at a
$\chi^2$ of 0.21 that is lower than the accepted model's 5.79.
(b)~The exact forward solution (dashed curves) evaluated at exactly the
parameters that trajectory reports, against the same data, with the
hard-mode M4 solution (gray dotted) for reference: $\chi^2 = 16.6$. The trajectory absorbed the
misfit as slack in the governing equation, so the apparent quality of
the fit in (a) is not a property of the recovered parameters. The
factor of roughly 80 between the two is the diagnostic; the hard mode
returns 5.79 by either route and passes by construction.}
\label{fig:f1}
\end{figure}

The soft-mode inversion behaves differently. Trained on the same data,
the joint field-and-parameter optimization settles into an equilibrium
at which the data loss is essentially zero while the governing equation
along the recovered trajectory is not satisfied to the precision the
low data loss suggests. Re-solving the recovered parameters through the
exact forward solution exposes the inconsistency directly: the
trajectory's own $\chi^2$ is 0.21, but the same parameters evaluated
through the exact solution give $\chi^2 = 16.6$
Fig.~\ref{fig:f1} and Table~S6), a factor of roughly 80
apart.
Increasing the fixed weight on the physics residual does not resolve
this, because the adaptive loss aggregator renormalizes relative term
weights during training; an independent reweighting toward the data
term reproduces the same signature (trajectory 0.67 versus closed-form
15.6). The specific values vary from training run to run, but the
one-to-two-orders-of-magnitude signature is stable (SI, Section~S4.4). We term
this failure mode F-1 and attribute it to the gradient geometry near
the data manifold: once the data loss is near zero, the physics-residual
gradient is too small to pull the trajectory off that manifold. This
mechanism, and the expectation that dense data would prevent the
imbalance, are offered as a working explanation consistent with the
observations rather than as a demonstrated result.

The practical fix is the hard-mode construction, available for the
zero-dimensional kinetics but not, in general, for the spatial
extension. For soft-mode inversions we therefore apply, and propose for
reporting alongside any physics-informed inverse result, a
self-consistency check: re-solve the recovered parameters through an
independent physics-exact forward solve (closed form, Runge--Kutta, or
Newton iteration) and report the resulting $\chi^2$ next to the
trajectory $\chi^2$. In our experience this is the only test among
those we tried that reliably distinguishes a trustworthy soft-mode fit
from an F-1 failure without requiring a hard-mode alternative to exist;
its acceptance rule is a stated heuristic (SI, Section~S4.4).

\section{Discussion}\label{sec:discussion}

The discussion falls into two halves. The first asks what the identified
parameters mean: what the empirical exponents become under a
mechanistic reading, what the Pilling--Bedworth ratio can be compared
against, and what the nickel zone implies about transport at the
receding interface. The second bounds how far those readings can be
taken: what is and is not attributable to the alloy, what the field
strength requires from outside the data, whether the closure that
defines it survives a self-consistent solve, and what the calibration
environment leaves untested.

The mechanistic model supports a physical reading of Veile et al.'s
fitted exponents that the power-law description cannot offer. The
near-parabolic chromium exponent ($n \approx 0.50$) is the signature of
a barrier layer growing under near-zero dissolution in ultrapure water;
the sub-parabolic iron exponent ($n \approx 0.22$) reflects deposition
constrained by constant-volume coupling to the barrier layer from a
substantial initial precipitate population, rather than independent
iron-layer kinetics. This reading is available regardless of fit
quality, and it is the coupling behind it, not $\chi^2$, that separates
the two descriptions' ten-year predictions by a factor of 3.5 to 6.3.

That coupling is carried by a single parameter, so what its value can
be checked against decides how much weight it carries. The identified
$\mathrm{PBR}_\text{eff} = 1.096$ is one of the two
quantities the exposure-time data constrain only weakly, so what it does
and does not agree with needs stating exactly. An earlier
version of this analysis placed a prior on it centered on 1.05 and
described that center as the stoichiometric spinel/magnetite value. It is
not. The constant-volume closure that yields 1.05 requires the barrier
layer to draw approximately 30~at\% chromium from an alloy containing
18.8~at\%, which the alloy cannot supply. A chromium mass balance on
X6CrNiNb18-10 with an FeCr$_2$O$_4$ barrier gives
$\mathrm{PBR}_\text{eff} = 2.05$ if every iron atom not retained in the
nickel-enriched zone reports to the outer layer, and 2.16 if the nickel routing fraction of Section~\ref{sec:nickel} is used in place of complete rejection. That second figure inherits the caveat of Section~\ref{sec:nickel}: the routing fraction comes from a closure the shape test rejects, so it is admissible here only as a bound on the range, never as an estimate. The first figure is the natural point of comparison, but, as the rest of this section shows, it is neither unique nor a bound.

The likelihood puts the parameter at 1.10, a factor of 1.9 below that
figure. That gap is not a conflict, for two independent reasons, and
neither of them requires the fit to be wrong.

The first is that 2.05 is not a bound but one value of a
phase-dependent quantity. It follows from assigning the barrier the
composition FeCr$_2$O$_4$, and the photoelectron spectroscopy admits
three further chromium-rich spinels. Carrying the same mass balance
through each of them (using lattice-parameter molar volumes, the
nominal alloy composition, and, in one variant, the nickel routing
fraction identified in Section~\ref{sec:spatial} rather than an
assumption of complete nickel rejection) gives production ratios, under complete nickel
rejection, of 2.05 (FeCr$_2$O$_4$), 2.50 (NiCr$_2$O$_4$), 3.64
(Cr$_2$O$_3$) and 0.52 (Fe$_2$CrO$_4$); routing nickel at the identified
fraction instead moves these to 2.16, 2.50, 3.79 and 0.57. Taken over
both routing assumptions the stoichiometrically admissible range is
$[0.52, 3.79]$, and the identified 1.10 lies inside it. Two of the eight values lie below 2, so the FeCr$_2$O$_4$ assignment is one reading
of the stoichiometry rather than the reading.

The second reason is that, at any fixed phase assignment, the two
numbers are not the same quantity. A mass balance fixes the rate at
which barrier growth \emph{produces} outer-layer oxide. The accepted
model pins $C_x = 0$, so its $\mathrm{PBR}_\text{eff}$ measures net
\emph{accumulation} instead, and the two differ by whatever oxide leaves
for the coolant. Freeing $C_x$ separates them, and doing so shows the
data cannot: imposing the FeCr$_2$O$_4$ value of 2.05 changes $\chi^2$
from 5.79 to 5.67, a \emph{decrease} of 0.12
(Table~\ref{tab:pbrclosure}), paid for by an outer-layer loss rate of
0.24~nm/h and a smaller initial precipitate population, 94~nm against
the 120~nm the same $C_x$-free family returns at the fitted ratio
(Table~\ref{tab:pbrclosure}). Profiling $\mathrm{PBR}_\text{eff}$ from 0.5 to 8 in this family
moves $\chi^2$ by 0.89 in total, so the $\Delta\chi^2 = 1$ interval is
the entire scanned range. The reason is structural rather than
statistical: with $L_\text{bl}(t)$ fixed, the outer-layer solution
$L_\text{ol}(t) = L_\text{ol,0} + \mathrm{PBR}_\text{eff}\,
(L_\text{bl} - L_0) + C_x t$, with $C_x \le 0$ as in
Eq.~\eqref{eq:odeol}, carries three parameters against exactly
three outer-layer observations, so it is saturated and no combination is
preferred.

The correct statement is therefore not that the identification disagrees
with stoichiometry, but that three exposure times cannot separate oxide
production from oxide loss. This also explains an observation that would
otherwise look like a loose end: the M1 and M2 variants, which lack the
initial precipitate population, reach $\mathrm{PBR}_\text{eff} = 2.43$;
they are simply at a different point on the same flat direction. The
resolution makes one falsifiable prediction. If the barrier is
FeCr$_2$O$_4$ and stoichiometry holds, the outer layer must lose about
9~$\mu$g of iron per square decimeter per hour, a magnitude
consistent with the corrosion-product release reported for stainless
steels in high-temperature water~\citep{lister1987release}. Nothing
measured here demonstrates it: this dataset contains no coolant
chemistry, and the outer layer's scan-to-scan scatter is so large that
it carries only 5\% of the fit statistic (Table~\ref{tab:leverage}). A
coolant mass balance, or outer-layer thicknesses at more exposure times,
would decide it.

\begin{table}[htbp]
\centering
\caption{Imposing a stoichiometric $\mathrm{PBR}_\text{eff}$ once the
outer-layer loss term $C_x$ is free. Every row is a refit of that
$C_x$-free family at the ratio shown, the first row included, so the
first row is not the accepted model but the same family evaluated at the
accepted model's ratio; it returns $C_x \approx 0$, recovering the
accepted fit. Every larger ratio is reached at no cost in $\chi^2$, paid
for by outer-layer loss and a smaller initial precipitate population.
The accepted model has $\chi^2 = 5.79$, $C_x = 0$ and
$L_\text{ol,0} = 115.6$~nm.}
\label{tab:pbrclosure}
\begin{tabular}{lcccc}
\toprule
$\mathrm{PBR}_\text{eff}$ imposed & $\chi^2$ & $\Delta\chi^2$ &
$C_x$ (nm/h) & $L_\text{ol,0}$ (nm) \\
\midrule
1.10 \; (fitted value imposed)    & 5.784 & $-0.007$ & $-0.025$ & 120.2 \\
2.05 \; (FeCr$_2$O$_4$)           & 5.671 & $-0.120$ & $-0.241$ & 93.6 \\
2.16 \; (FeCr$_2$O$_4$, Ni routed) & 5.661 & $-0.131$ & $-0.264$ & 90.7 \\
2.43 \; (M1/M2 optimum)           & 5.635 & $-0.157$ & $-0.327$ & 83.0 \\
\bottomrule
\end{tabular}
\end{table}

The third feature the identified kinetics speak to is the one they do
not describe. The order-of-magnitude mobility argument of
Section~\ref{sec:nickel}, which depends on no fitted value and so
survives the rejection of the closure that produced it, places whatever
transports nickel at the receding metal/barrier-layer interface well
above lattice diffusion: it must be short-circuit or defect-flux
mediated. That is also the class of mechanism whose mobility would decay as the
interface structure coarsens, which is the natural explanation for the
flat measured zone widths that defeat the constant-mobility model
(Section~\ref{sec:nickel}). We stop there: distinguishing candidate
short-circuit mechanisms requires exposure-time resolution this dataset
does not have, and the constant-mobility fit is rejected by the data in
shape, so neither its incorporation ratio nor its nested-model
comparison against a re-precipitation-only outer-layer supply picture
can be offered as evidence either way.

Those three readings are as far as the identified parameters can be
taken forward; the rest of this section bounds how far they can be taken
at all. One bound concerns the alloy: whether any identified
parameter registers the niobium stabilization that is the reason this
grade is used in reactor internals. The reaction network fitted here
contains nothing niobium-specific, and none of the five identified
parameters can be attributed to Nb stabilization on the evidence
presented. Being explicit about what that means matters, since the
alloy's stabilization is the reason it is used in reactor internals. The
identification is specific to AISI~347 in the sense that the numbers
come from measurements on AISI~347, not in the sense that the model
structure encodes Nb; every parameter is an apparent quantity for an
18Cr--10Ni austenitic matrix under this environment, and the same
structure fitted to 304 or 316 data in the same environment would differ
only through the fitted values. Distinguishing a genuine stabilization
effect would require the comparison this dataset does not contain:
matched exposures of stabilized and unstabilized grades, in which NbC
precipitates could be shown to alter local chromium availability at the
metal/barrier-layer interface, to nucleate or pin the inner spinel, or
to change the initial film that $L_0$ and $L_\text{ol,0}$ parametrize.
Two of the identified quantities are the natural place such an effect
would appear: the growth prefactor $A_\text{bl}$, through chromium
supply to the growth reaction, and the initial-condition parameters,
through surface and precipitate structure. Both are reported here with
intervals wide enough that a moderate stabilization effect would be
invisible. The claim we make is therefore that this is the first
mechanistic identification for this alloy in this environment, not that
the mechanism identified is peculiar to the stabilized grade.

A second bound attaches to the field strength. Of the five apparent
parameters, the field exponent has the most direct physical reading and
demands the most from outside the data to obtain it. Converting the identified field exponent through
Eq.~\eqref{eq:recover} requires the growth-reaction transfer
coefficient $\alpha_3$, which single-temperature thickness data cannot
supply. At the reference value $\alpha_3 = 0.12$ fit by Li et
al.~\citep{li2020scwpdm} for a ferritic steel in supercritical water,
$\varepsilon_f = 1.7\times10^4$~V/cm; over the physically defensible
range $\alpha_3 \in [0.1, 0.5]$ the derived field spans
$4.1\times10^3$ to $2.1\times10^4$~V/cm. The implied potential drop
across the 480-h barrier layer, $\varepsilon_f L$, is correspondingly
0.06--0.28~V; mixed-potential estimates of the corrosion potential in
oxygenated 288$^\circ$C BWR water are of order 0 to
$+150$~mV on the standard hydrogen electrode (SHE)
scale at 0.2--0.4~ppm
O$_2$~\citep{macdonald1992hwc,lin1996ecp}, which favors the lower end
of the derived range and is one more reason to treat $\varepsilon_f$
here as an order-of-magnitude quantity. Within that reading, the value
sits plausibly between the $\sim10^2$~V/cm of supercritical-water PDM
fits at 500$^\circ$C~\citep{li2020scwpdm} and the $\sim10^6$~V/cm of
room-temperature passive films~\citep{cabrera1949oxidation}; we offer
this ordering as a weak plausibility remark only, since the three
anchors span different alloy classes and film chemistries, and the
nearest-temperature precedents (PDM and mixed-conduction analyses of
304/316 in 280--320$^\circ$C
water~\citep{ss316l_pwr_pdm,bojinov_mcm,ss316ln_pdm2024}) do not report
directly comparable apparent field strengths.

The identified kinetics sit within the high-temperature-water oxide
corpus, tabulated against related systems in the SI (Table~S8). The inner-layer thickness
trajectory recovered here (41~nm at 72~h to 135~nm at 480~h) is of the
same order as inner-layer thicknesses reported for 304/316 stainless
steels in oxygenated 288$^\circ$C water at comparable
times~\citep{kuang2010oxidation,kim1999oxide}, and the duplex
morphology, inner-layer spinel chemistry, and near-parabolic inner
growth match the established picture for austenitic steels in this
environment~\citep{robertson1991mechanism,stellwag1998mechanism,ziemniak2002corrosion}.
The ten-year prediction of 535~nm has no direct validation data for
AISI~347; we are not aware of published long-term harvested-component
oxide thicknesses for this alloy in BWR conditions, and the prediction
should be read as a testable statement, not a validated one.

\begin{figure}[htbp]
\centering
\includegraphics[width=\textwidth]{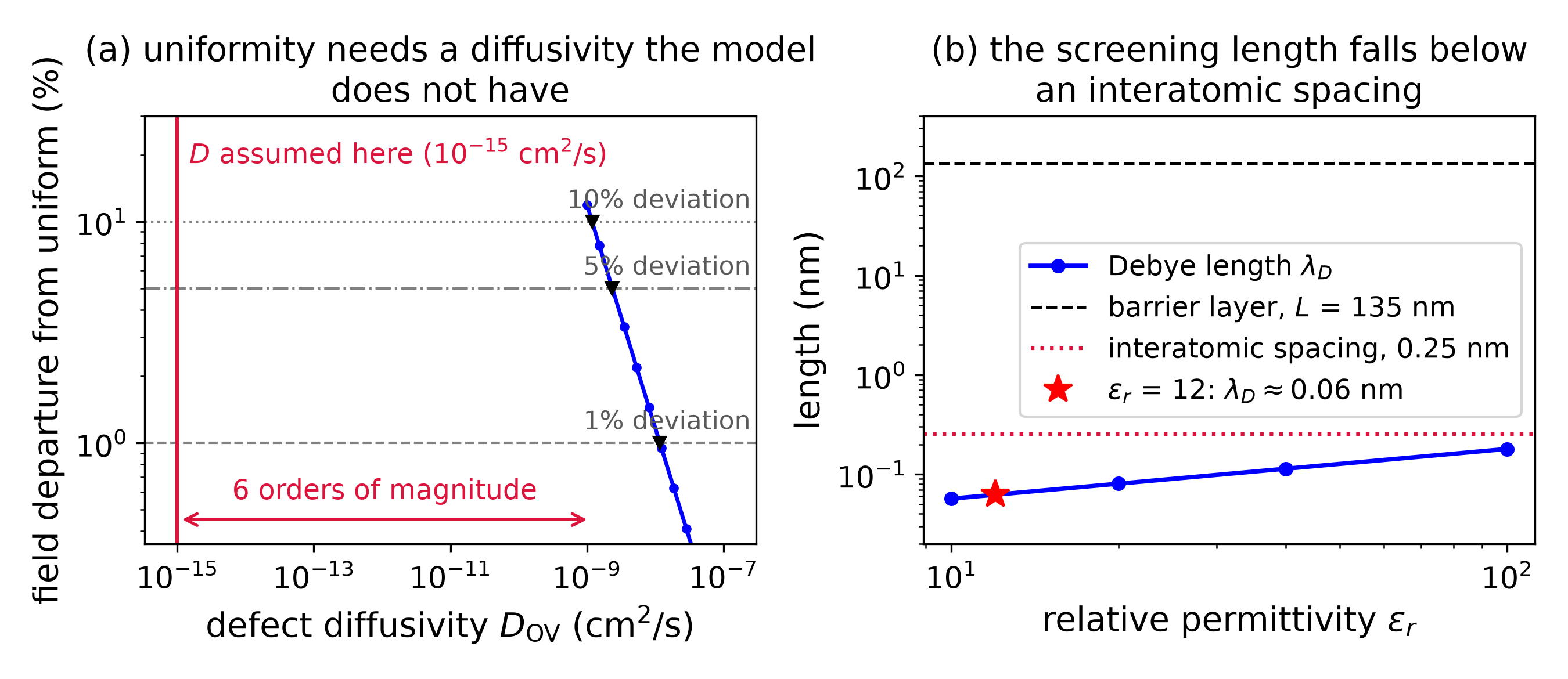}
\caption{The constant-field closure is not self-consistent at the
transport constants the model assumes. (a)~Maximum departure of the
self-consistently solved field from uniform, against the oxygen-vacancy
diffusivity that would be needed to hold it there; markers locate the
1\%, 5\% and 10\% thresholds. The diffusivity actually assumed lies six
orders of magnitude below all three. (b)~The Debye length against its
two yardsticks: the barrier layer it must screen across, and an
interatomic spacing. At the permittivity used here it is approximately
0.06~nm, several times smaller than the smaller of the two, which is
what makes the closure untenable rather than merely approximate.}
\label{fig:closure}
\end{figure}

Those field strengths rest on an assumption
the identification itself never tests. Every number derived from $b_3$
through
$b_3 = -\alpha_3\chi\gamma\varepsilon_f$ inherits the PDM's constant-field
closure, and that closure can be tested directly by solving Poisson's
equation together with the defect transport instead of imposing the
field on it. We have carried out that solve, verifying it by reduction:
with the space charge switched off it reproduces the analytic
constant-field solution to $1\times10^{-15}$. Switched on, it does not
support the closure (Fig.~\ref{fig:closure}). At the identified kinetics and the transport
constants assumed in Section~\ref{sec:spatial}, the dimensionless
space-charge parameter is of order $10^{4}$--$10^{5}$ and the
corresponding Debye length is approximately 0.06~nm, several times
smaller than an interatomic spacing; holding the field within 10\% of
uniform would require defect diffusivities near
$10^{-9}$--$10^{-10}$~cm$^2$/s, some six orders of magnitude
above the assumed values, or equivalently defect concentrations below
about $10^{14}$~cm$^{-3}$. The implication is not that the field
distribution is merely non-uniform but that a two-species charged-defect
description cannot produce a constant field at the concentrations the
model itself implies: charge compensation by electronic carriers, which
the PDM does not carry, is required. This leaves the fitted apparent
parameters and every prediction drawn from them untouched, since the
identification is of $b_3$ and not of $\varepsilon_f$. What it bounds is
the interpretation: $\varepsilon_f$ should be read as an apparent
field-related grouping, and the numerical field strengths
quoted just above carry that caveat.

A third bound is the calibration dataset itself, which imposes four
conditions on everything above, each a property of the measurements
rather than of the analysis. First, the kinetics
rest on a single temperature and a single dissolved-oxygen level, so the
operating-envelope results of Section~\ref{sec:predictions} are forward
projections under sourced assumptions, not validation against
multi-condition data, which does not exist for this alloy. Their
temperature channel additionally assumes a single rate-determining step
with a temperature-independent field strength; the activation-energy
scan bounds the parameter of that channel but not its structure. Their
oxygen channel is an ideal-Nernst shift of the growth prefactor, and
measured corrosion-potential--oxygen curves in BWR-relevant chemistry
are steeper than Nernstian in the parts-per-billion transition
region~\citep{lin1996ecp}, so the mapped oxygen sensitivity is a lower
bound on the true response. Second, the exposures are unirradiated and
static, while radiolysis products, H$_2$O$_2$ in particular, are known
to alter oxide structure and thickness on stainless steels relative to
O$_2$ at comparable corrosion potential~\citep{kim1999oxide}, and flow
affects outer-layer growth and dissolution through solubility and mass
transfer~\citep{lister1987release}; neither channel is in the model.
Third, surface finish conditions the initial film and the early
precipitate population, the parameters $L_0$ and $L_\text{ol,0}$ here,
and mechanical surface treatments are documented to change oxide
properties on 304L in simulated BWR and PWR
water~\citep{surface_treatment_304l}, so the identified $L_\text{ol,0} =
116$~nm is specific to Veile et al.'s coupon preparation. Fourth,
EDX-derived layer thicknesses are not total oxide mass, so the
composition comparison of Section~\ref{sec:spatial} is a shape and
plateau-composition cross-check rather than a mass-balance validation,
which matters for the iron-release prediction made above,
since that prediction is built on
thicknesses and would be decided by a coolant mass balance.

A bound of a different kind attaches to the inversion machinery. Nothing
in this paper's physical conclusions requires the neural solver: the
kinetics have a closed form, the steady transport problem has an
analytic reference, and the classical methods used to check the neural
results would have produced the same numbers, faster, and in the spatial
case to eleven more digits. That is a deliberate property of the test
case rather than an oversight, since a validation is possible only where
an exact reference exists, but it bounds the claim to this: the
framework is trustworthy on real sparse data once the self-consistency
check of Section~\ref{sec:pinn} is applied, not that it was needed here.
The problems that would need it are visible from this study. Two lie
inside the present model: solving the potential distribution
self-consistently with defect transport instead of imposing the
constant-field assumption, and inverting the moving-boundary transient
for kinetic parameters and defect fields together, where the classical
route requires a remeshed forward solve inside every optimizer
iteration. A third is sharper: the nickel zone rejects every
constant-mobility description these data can test, so the quantity
actually wanted is a mobility \emph{function} of unknown form, which a
neural representation can carry as an unknown closure constrained by the
transport equation while a classical solver must first commit to a
parametric guess.

\section{Conclusions}\label{sec:conclusions}

What remains after all four bounds is the paper's recurring finding.
Every identifiability limitation encountered here (the zero-dimensional
profile analysis, the residual-leverage imbalance, the spatial gate and
the nickel-closure gate) traces to too few independent exposure times.
The convergence of four unrelated analyses on one cause is the
methodological result, and the remedy it implies is concrete enough to
cost: a designed exposure near 3000~h would separate the mechanistic
prediction from the power-law extrapolation at 2$\sigma$ against the full
error budget, and would sharpen every weakly constrained parameter at the
same time. For service-life assessment that is the one statement
this dataset supports that can be acted on, and it calls for an
experiment rather than a reanalysis.
Neither the cause nor the remedy is specific to this alloy or to the
point defect model; both apply to any sparse-data mechanistic
identification in which extrapolation beyond the calibration window is
the purpose of the fit.

\section*{Data availability}

The digitized Veile et al.\ layer-thickness data
(Table~\ref{tab:data}) are transcriptions of published figures,
released with the analysis repository below together with provenance
notes tracing each value to its source figure; this constitutes reuse of
published data requiring citation, not separate data-sharing permission.
A committed artifact stands behind every number reported here, and a
generated document maps each figure and table to the exact command that
produces it, so any reported value can be re-derived individually.

\section*{Code availability}

All code implementing the deterministic fits, physics-informed forward
and inverse solves, identifiability analyses, and spatial extension is
released as a standalone, installable Python package
(\texttt{htw-pdm}) at
\url{https://github.com/Feugmo-Group/x6crninb18-pdm} under the Apache
2.0 license, together with the configuration files and the analysis and
figure scripts. The classical fits, the mass balance and the spatial
solves depend only on NumPy, SciPy and Matplotlib; the neural inversions
additionally require the NSEM solver~\citep{feugmo2026nsem}, whose
spectral-element-network module
(\texttt{experimental.\allowbreak models.\allowbreak scen}) is
released with PhysicsNeMo, and the repository records the exact revision
used here.

\section*{Acknowledgments}

C.G.T.F. acknowledges the support of the Natural Sciences and
Engineering Research Council of Canada (NSERC)
[RGPIN-2024-03989]. This research was enabled in part by support
provided by SHARCNET and the Digital Research Alliance of Canada.

\section*{Author contributions}

\textbf{Conrard Giresse Tetsassi Feugmo:} Conceptualization,
Methodology, Software, Validation, Formal analysis, Investigation,
Writing -- original draft, Writing -- review \& editing, Visualization.

\section*{Competing interests}

The author declares no competing interests.

\bibliographystyle{unsrtnat}
\bibliography{references}

\end{document}